\documentclass[12pt]{iopart}
 \input{epsf.sty}
 \input{epsf.tex}
 \eqnobysec

\newcommand{\beq}{\begin{equation}}
\newcommand{\eeq}{\end{equation}}
\newcommand{\bea}{\begin{eqnarray}}
\newcommand{\eea}{\end{eqnarray}}

\def\square{\hbox to1pt{\hfill}\mathchoice\sqr84\sqr84
\sqr84\sqr84\hbox
       to1pt{\hfill}}
\def\sqr#1#2{{\vcenter{\hrule height.#2pt \hbox{\vrule width.#2pt
       height#1pt\kern#1pt \vrule width.#2pt} \hrule height.#2pt}}}

\def\laq{\raise 0.4ex\hbox{$<$}\kern -0.8em\lower 0.62
ex\hbox{$\sim$}}
\def\gaq{\raise 0.4ex\hbox{$>$}\kern -0.7em\lower 0.62
ex\hbox{$\sim$}}

\def \pa {\partial}
\def \ti {\tilde}
\def \se {{\prime\prime}}
\def \ra {\rightarrow}
\def \la {\lambda}
\def \La {\Lambda}
\def \Da {\Delta}
\def \b {\beta}
\def \a {\alpha}
\def \ap {\alpha^{\prime}}
\def \Ga {\Gamma}
\def \ga {\gamma}
\def \sg {\sigma}
\def \da {\delta}
\def \ep {\epsilon}
\def \r {\rho}
\def \om {\omega}
\def \Om {\Omega}
\def \noi {\noindent}

\def \hp {\dot{h}}
\def \hpp {\ddot{h}}

\def \fb {\overline \phi}
\def \rb {\overline \rho}
\def \pb {\overline p}
\def \fbp {\dot{\fb}}

\begin{document}


\begin{flushright}

BA-TH/99-345\\
hep-th/9907067
\end{flushright}

\vspace*{0.6truein}

{\large\bf\centering\ignorespaces
ELEMENTARY INTRODUCTION TO PRE-BIG BANG COSMOLOGY 
AND TO THE RELIC GRAVITON BACKGROUND
\vskip2.5pt}
{\dimen0=-\prevdepth \advance\dimen0 by23pt
\nointerlineskip \rm\centering
\vrule height\dimen0 width0pt\relax\ignorespaces

M. Gasperini 
\par}
{\small\it\centering\ignorespaces
Dipartimento di Fisica , Universit\`a di Bari, 
Via G. Amendola 173, 70126 Bari, Italy\\
and Istituto Nazionale di Fisica Nucleare, Sezione di Bari, Bari, Italy\\
\par}

\par
\bgroup
\leftskip=0.10753\textwidth \rightskip\leftskip
\dimen0=-\prevdepth \advance\dimen0 by17.5pt \nointerlineskip
\small\vrule width 0pt height\dimen0 \relax

\vspace*{0.6truein}

\centerline{\bf Abstract}

\noi
This is a contracted version of a series of lectures for graduate
and undergraduate students given at the {\sl ``VI Seminario Nazionale
di Fisica Teorica"} (Parma, September 1997), 
at  the Second Int. Conference {\sl
``Around VIRGO"}  (Pisa, September 1998), and at the Second
{\sl SIGRAV} School on  {\sl ``Gravitational Waves in Astrophysics,
Cosmology and String Theory"} (Center  ``A. Volta", Como, April 1999).
The aim is to provide an elementary, self-contained  introduction to
string cosmology and, in particular, to the background of relic cosmic
gravitons predicted in the context  of the so-called ``pre-big bang"
scenario.  No  special  preparation is required besides a basic
knowledge of general relativity and of standard (inflationary)
cosmology. All the essential computations are reported in full details
either in the main text or in the Appendices. For a  deeper and more
complete approach to the  pre-big bang scenario the interested reader is
referred to  the updated  collection of papers available at   
 {\tt http://www.to.infn.it/\~{}gasperin/}. 

\vspace{1.5cm}
\begin{center}
------------------------------  

\vspace{1.5cm}
To appear  in {\sl Proc. of the Second SIGRAV School on\\ 
``Gravitational
Waves  in Astrophysics, Comology and String Theory"}\\
Villa Olmo, Como, 19-24 April 1999 -- 
Eds. V. Gorini et al.\\
\end{center}

\thispagestyle{plain}
\par\egroup

\vfill

\maketitle

\setcounter{page}{1}

\title[Pre-big bang cosmology]{Elementary introduction to   pre-big
bang cosmology and to the relic graviton background}

\author{M. Gasperini\dag\ddag
\footnote[3]{E-mail: gasperini@ba.infn.it.}}

\address{\dag Dipartimento di Fisica , Universit\`a di Bari, 
Via G. Amendola 173, 70126 Bari, Italy}

\address{\ddag Istituto Nazionale di Fisica Nucleare,  Sezione di Bari,
Bari, Italy}

\begin{abstract}
This is a contracted version of a series of lectures for graduate
and undergraduate students given at the {\sl ``VI Seminario Nazionale
di Fisica Teorica"} (Parma, September 1997), 
at  the Second Int. Conference {\sl
``Around VIRGO"}  (Pisa, September 1998), and at the Second
{\sl SIGRAV} School on  {\sl ``Gravitational Waves in Astrophysics,
Cosmology and String Theory"} (Center  ``A. Volta", Como, April 1999).
The aim is to provide an elementary, self-contained  introduction to
string cosmology and, in particular, to the background of relic cosmic
gravitons predicted in the context  of the so-called ``pre-big bang"
scenario.  No  special  preparation is required besides a basic
knowledge of general relativity and of standard (inflationary)
cosmology. All the essential computations are reported in full details
either in the main text or in the Appendices. For a  deeper and more
complete approach to the  pre-big bang scenario the interested reader is
referred to  the updated  collection of papers available at   
 {\tt http://www.to.infn.it/\~{}gasperin/}. 

\end{abstract}

~~~~~~~~~~~~~Preprint  BA-TH/99-345 
~~~~~~~~~~~~~E-print Archives: hep-th/9907067


\section{Introduction}
\label{I}

The  purpose of these lectures is to provide an introduction 
to the background of relic gravitational waves expected in a string
cosmology context, and to discuss its main properties. To this purpose,
it seems  to be appropriate to include a short 
presentation of string cosmology,  in order to explain the basic ideas
underlying the so-called pre-big bang scenario, which is  
one of the most promising scenarios for the production of a
detectable graviton background of cosmological origin. 

After a short, qualitative presentation of the pre-big bang models, I will
concentrate on the details of the cosmic graviton spectrum: I will discuss
the theoretical predictions for different models, and I will compare the
predictions with existing phenomenological constraints,  and with the
expected sensitivities of the present gravity wave detectors.  A consistent
part of these lectures will thus be devoted to introduce  the basic notions
of cosmological perturbation theory,  which are required to compute the
graviton spectrum and to understand why the amplification of tensor
metric perturbations, at high frequency,  is more efficient in string
cosmology than in the standard inflationary context.

Let me start by noting that a qualitative, but effective
representation  of the main difference between string cosmology and
standard, inflationary cosmology can be obtained by plotting  the
curvature scale of the Universe versus time, as illustrated in Fig. 1. 

\begin{figure}
\centerline{\epsfxsize=8.0cm
\epsffile{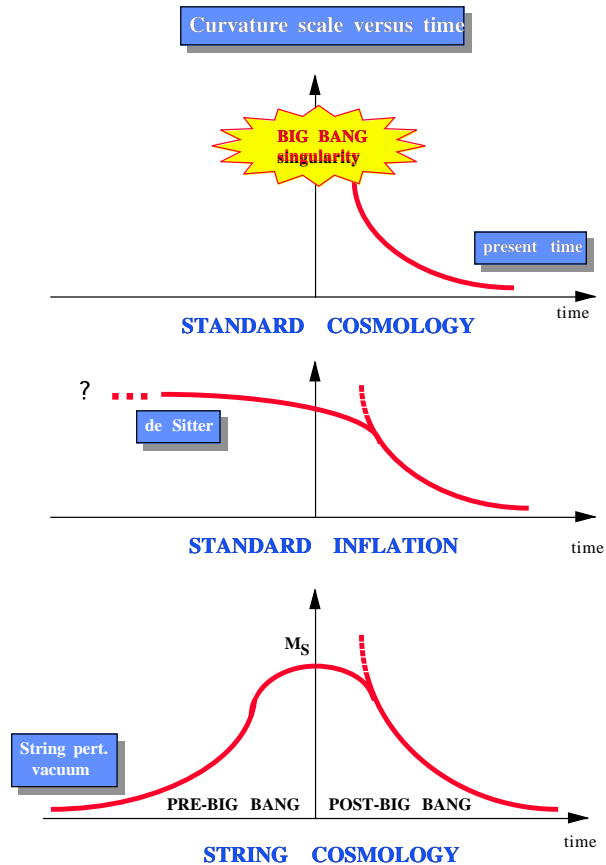}}
\caption{\sl Time evolution of the curvature scale in the standard
cosmological scenario, in the conventional inflationary scenario, and in
the string cosmology scenario.}
\label{fig1}
\end{figure}

According to the cosmological solutions of the so-called ``standard" 
scenario \cite{1}, the spacetime curvature decreases in time.
As we go back intime the curvature  grows monotonically, and blows
up at the initial ``big bang" singularity, as illustrated in the top part of
Fig. 1 (a similar plot, in the standard scenario, also describes the
behaviour of the temperature and of the energy density of the
gravitational sources).

According to the standard inflationary scenario \cite{2}, on the
contrary, the Universe in the past is expected to enter a de Sitter, or
``almost" de Sitter phase, during which the curvature tends to stay
frozen at a nearly constant value. From a classical point of view,
however, this scenario has a problem, since a phase of expansion at
constant curvature cannot be extended back in time for ever \cite{3}, for 
reasons of geodesic completeness.  This point was clearly stressed also
in  Alan Guth's recent survey of inflationary cosmology 
\cite{4}:

\bigskip 

{\sl ``... Nevertheless, since inflation appears to be eternal only into
the future,  but not to the past, an important question remains open.
How did  all start? Although eternal inflation pushes this question far
into the  past, and well beyond the range of observational tests, the
question  does not disappear."}

\bigskip 

A possible anwer to this question, in a quantum cosmology context, is
that the Universe  emerges in a  de Sitter state ``from nothing" 
\cite{5} (or from some unspecified ``vacuum"), though a process of
quantum tunnelling. I will not discuss the quantum approach in these
lectures, but let me note that  the computation of the transition probability
requires an appropriate choice of  the boundary conditions \cite{6}, which
in the context of standard inflation are imposed {\sl ``ad hoc"} when
the Universe is in an unknown state,  deeply inside the non-perturbative, 
quantum gravity regime.  In a string cosmology context, on the contrary,
the initial conditions are referrred asymptotically to a low-energy,
classical state which is known, and well controlled by the low-energy
string effective action \cite{7}.

From a classical point of view, however, the answer to the
above question -- what happens to the Universe before the phase 
of constant curvature, which cannot last for ever -- is very simple, as
we are left only with two possibilities. Either the curvature starts
growing again, at some point in the past (but in this case the
singularity problems remains, it is simply shifted back in time), or
the curvature starts decreasing. 

In this second case we are just led to the string cosmology scenario,
illustrated in the bottom part of Fig. 1. String theory suggests indeed
for the curvature a specular behaviour (or better a ``dual" behaviour,
as we shall see in a moment) around the time axis. As we go back in
time the curvature grows, reaches a maximum controlled by the string
scale, and then decreases towards a state which is asymptotically
flat and with negligible interactions (vanishing coupling constants),
the so-called ``string perturbative vacuum".   In this 
scenario the phase of high, but finite (nearly Planckian) curvature is
what replaces the big bang singularity of  the standard scenario. It
comes thus natural, in a string cosmology context, to call ``pre-big
bang" \cite{8} the initial phase with growing curvature, in contrast to
the subsequent,  standard, ``post-big bang " phase with decreasing
curvature. 

At this point, a number of questions may arise naturally. In
particular:
               
\begin{itemize}

\item{}{\bf \underline{Motivations}}: why such a cosmological
scenario, characterized by a ``bell-like" shape of the curvature,
seems to emerge in a string cosmology context and not, for
istance, in the context of standard cosmology based on the Einstein
equations?

 \item{}{\bf \underline{Kinematics}}: in spite of the differences, is the
kinematic of the pre-big bang phase still appropriate to solve the well
known problems (horizon, flatness ...) of the standard scenario? After all,
we do not want to loose the main achievements of the conventional
inflationary models.

\item{}{\bf \underline{Phenomenology}}: are there phenomenological
consequence that can discriminate between string cosmology models
and other inflationary models? and are such effects observable, at
least in principle?

\end{itemize}

In the following sections I will present a quick discussion of the
three points listed above.

\section{Motivations: duality symmetry}
\label{II}

There are various motivations, in the context of string theory,
suggesting a cosmological scenario like that illustrated in Fig. 1. 
All the motivations are however related, more or less directly, to an
important property of  string theory, the duality symmetry of the
effective action. 

To illustrate this point, let me start by recalling that in general
relativity the solutions of the standard Einstein action, 
\beq
S= -{1\over 2 \la_p^{d-1}}\int d^{d+1}x \sqrt{|g|}~ R
\label{21}
\eeq
($d$ is the number of spatial dimensions, and $\la_p=M_p^{-1}$ is the
Planck length scale), are invariant under ``time-reversal"
transformations. Consider, for instance, a homogeneous and isotropic
solution of the cosmological equations, represented by a scale factor
$a(t)$: 
\beq
ds^2= dt^2 - a^2(t) dx_i^2.
\label{22}
\eeq
If $a(t)$ is a solution, then also $a(-t)$ is a solution. On the other
hand, when $t$ goes into $-t$, the Hubble parameter $H=\dot a/a$
changes sign,
\beq
a(t) \rightarrow a(-t), ~~~~~~~~~~~~~~~~
H=\dot a/a \rightarrow -H. 
\label{23}
\eeq
To any standard cosmological solution $H(t)$, describing
decelerated expansion and decreasing curvature ($H>0$, $\dot H <0$), 
is thus associated a ``reflected" solution, $H(-t)$, describing a
contracting Universe  because $H$ is negative. 

This is the situation in general relativity. In string theory the action, in
addition to the metric, contains at least another fundamental field, the
scalar dilaton $\phi$. At the tree-level, namely to lowest order in the
string coupling and in the higher-derivatives ($\ap$) string corrections, the
effective action which guarantees the absence of conformal anomalies for
the motion of strings in curved backgrounds (see the Apppendix A) 
can be written as:  
\beq 
S= -{1\over 2\la_s^{d-1}} \int d^{d+1}x \sqrt{|g|}~
e^{-\phi}\left[R+ \left(\pa_\mu\phi\right)^2\right]
\label{24}
\eeq
( $\la_s=M_s^{-1}$ is the fundamental string length scale; see the
Appendix B for notations and sign conventions).  In addition to the
invariance under time-reversal, the above action is also invariant
under the ``dual" inversion of the scale factor, accompanied by an
appropriate transformation of the dilaton (see \cite{9} and the first paper
of Ref. \cite{8}).  More precisely, if  $a(t)$ is
a solution for the cosmological background (\ref{22}) , then 
$a^{-1}(t)$ is also a solution, provided the dilaton transforms as:
\beq
a \rightarrow \tilde a = a^{-1}, ~~~~~~~~~~~~~~~~
\phi \rightarrow \tilde \phi = \phi- 2 d \ln a
\label{25}
\eeq
(this transformation implements a particular case of $T$-duality
symmetry, usually called ``scale factor duality", see the 
Appendix B). 

\begin{figure}
\centerline{\epsfxsize=8.0cm
\epsffile{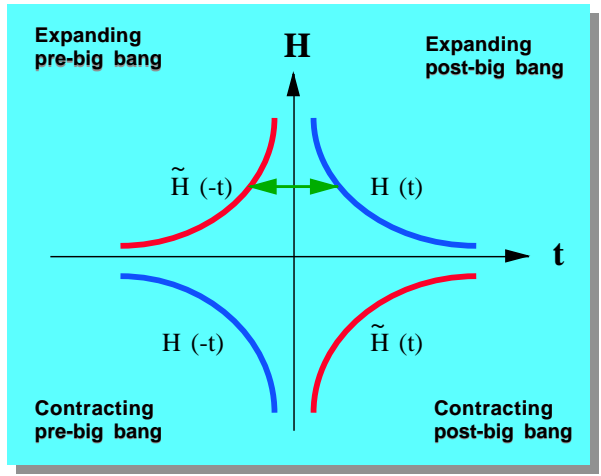}}
\caption{\sl The four branches of a low-energy string cosmology
background.} 
\label{fig2}
\end{figure}

When $a$ goes into $a^{-1}$, the Hubble parameter $H$ again goes into
$-H$ so that, to each one of the two solutions related by time
reversal, $H(t)$ and $H(-t)$, is associated a dual solution, 
$\tilde H(t)$ and $\tilde H(-t)$, respectively (see Fig. 2). 
The space of solutions is thus reacher in a string cosmology context.
Indeed, because of the combined invariance under the transformations
(\ref{23}) and (\ref{25}), a cosmological solution has in general four
branches: two branches describe expansion (positive $H$), two
branches describe contraction (negative $H$). Also, as illustrated in
Fig. 2, for two branches the curvature scale ($\sim H^2$) grows  in
time, with a typical ``pre-big bang" behaviour, while for the other
two branches the  curvature scale decreases,  with a typical
``post-big bang" behaviour. 

It follows, in this context,  that to any given decelerated 
expanding solution, $H(t)>0$, with decreasing curvature, $\dot H(t)<0$ 
(typical of the standard cosmological scenario), is always associated a
``dual partner" describing accelerated expansion, $\tilde H(-t)>0$, 
and growing curvature, $\dot{\tilde H}(-t)>0$. This doubling of
solutions has no analogue  in the
context of the Einstein cosmology, where there is no dilaton, and  the
duality symmetry cannot be implemented. 

It should be stressed, before proceeding further, that the duality
symmetry is not restricted to the case of homogeneous and isotropic 
backgrounds like (\ref{22}),  but is expected to be a general property
of the solutions of the string effective action (possibly valid at all
orders \cite{10}, with the appropriate generalizations).  The inversion of
the scale factor, in particular,  and the associated transformation
(\ref{25}), is only a special case of a more general, $O(d,d)$ symmetry of
the string effective action, which is manifest already at the lowest order. 
In fact, the tree-level action in general contains, besides the metric and
the dilaton, also a second rank antisymmetric tensor $B_{\mu\nu}$, the
so-called Kalb-Ramond ``universal" axion:   \beq
 S= -{1\over 2\la_s^{d-1}}\int d^{d+1}x \sqrt{|g|}~ e^{-\phi}\left[R+
\left(\pa_\mu\phi\right)^2- {1\over 12}
\left(\pa_{[\mu}B_{\nu\a]}\right)^2\right] .
\label{26}
\eeq
Given a background, even anisotropic, but with $d$ abelian isometries,
this action is invariant under a global, pseudo-orthogonal group of
$O(d,d)$ transformations which mix in a non-trivial way the
components of the metric and of the antisymmetric tensor, leaving
invariant the so-called ``shifted" dilaton $\fb$:  
\beq
\fb= \phi -\ln \sqrt{|{\rm det} g_{ij}|}.
\label{27}
\eeq

In the particular, ``cosmological" case in which we are interested in, the
$d$ isometries  correspond to spatial traslations (namely, we are in the
case of a  homogeneous, Bianchi I type metric background). For this
background, the action (\ref{26}) can be rewritten in terms of the
$2d\times 2d$ symmetric matrix $M$, defined by the spatial components
of the metric, $g_{ij}$, and of the axion, $B_{ij}$, as:  
\beq
B\equiv B_{ij}, ~~~~~~~ G\equiv g_{ij},  ~~~~~~~
M=\pmatrix{G^{-1} & -G^{-1}B \cr
BG^{-1} & G-BG^{-1}B \cr}.
\label{28}
\eeq
In the cosmic time gauge, the action takes the form (see the
Appendix B) 
\beq
S=-{\la_s \over 2}\int dt~ e^{-\fb}\left[(\dot{\fb})^2+{1\over 8}
{\rm Tr}~ \dot M (M^{-1}) \dot{}~\right], 
\label{29}
\eeq
and is manifestly invariant under the set of global
transformations   \cite{11}:
\beq
\fb \ra \fb, ~~~~~~~~ M \ra \La^{T} M \La, ~~~~~~~~
\La^T\eta \La =\eta, ~~~~~~~~ \eta =
\pmatrix{0 & I \cr I & 0 \cr}, 
\label{210}
\eeq
where $I$ is the d-dimensional unit matrix, and $\eta$ is the $O(d,d)$
metric in off-diagonal form. The transformation (\ref{25}),
representing scale factor duality,  is now reproduced as a particular
case of (\ref{210}) with the trivial $O(d,d)$ matrix $\La =\eta$, and for
an isotropic background with $B_{\mu\nu}=0$. 

This $O(d,d)$ symmetry holds even in the presence of matter sources,
provided they transform according to the string equations of motion
in the given background \cite{12}. In the perfect fluid approximation,
for instance, the inversion of the scale factor corresponds to a
reflection of the equation of state, which preserves however the
``shifted" energy $\rb= \r |{\rm det} g_{ij}|^{1/2}$:
\beq
a \rightarrow \tilde a = a^{-1}, ~~~~~~~~
\fb \rightarrow \fb, ~~~~~~~~ p/\r \ra -p/\r, ~~~~~~~~
\rb \ra \rb
\label{211}
\eeq

A detailed discussion of the duality symmetry is outside the purpose
of these lectures. What is important, in our context, is the
simultaneous presence of duality and time-reversal symmetry: by
combining these two symmetries, in fact, it is possible in principle to
obtain cosmological solutions of the  ``self-dual" type,
characterized by the conditions
\beq
a(t) = a^{-1}(-t), ~~~~~~~~~~~~~~~~ \fb (t) = \fb (-t).
\label{212}
\eeq
They are important, as they connect in a smooth way the phase of
growing and decreasing curvature, and  also describe a smooth
evolution from the string perturbative vacuum (i.e. the asymptotic
no-interaction state in which $\phi \ra -\infty$ and the string
coupling is vanishing, $g_s= \exp(\phi/2) \ra 0$), to the present
cosmological phase in which the dilaton is frozen, with an 
expectation value \cite{13} $\langle g_s \rangle= M_s/M_p \sim 0.3$
--   $0.03$ (see Fig. 3).

The explicit occurrence of self-dual solutions and, more generally, of
solutions describing a complete and smooth transition between the
phase of pre- and post-big bang evolution, seems to require in
general the presence of higher order (higher loop and/or higher
derivative) corrections to the string effective action \cite{14} (see
however \cite {15,16}). So, in order to give only a simple example of
combined $\{$duality $\bigoplus$ time-reversal$\}$ transformation, let me
consider here the low-energy, asymptotic regimes,  which are well
described by the lowest order effective action.

\begin{figure}
\centerline{\epsfxsize=8.0cm
\epsffile{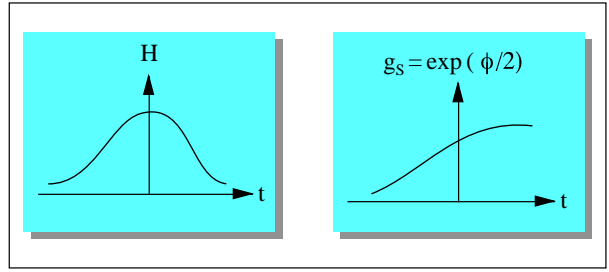}}
\caption{\sl Time evolution of the curvature scale $H$ and of the string 
coupling $g_s= \exp(\phi/2) \simeq M_s/M_p$, for a typical self-dual
solution of the string cosmology equations.}
\label{fig3}
\end{figure}

By adding matter sources, in the perfect fluid form, to the
action (\ref{24}), the string cosmology equations for a $d=3$,
homogeneous, isotropic and conformally flat background can be
written as (see Appendix C, eqs. (\ref{c14}), (\ref{c16}), (\ref{c13}),
respectively):  
\bea
&&
\dot \phi^2  -6 H\dot\phi+6 H^2=e^\phi \r,\nonumber\\
&&
\dot H -H \dot\phi +3H^2 ={1\over 2} e^\phi p, \nonumber\\
&&
2  \ddot \phi +6 H\dot\phi -\dot \phi^2 - 6\dot H-12 H^2=0.
\label {213}
\eea
For  $p=\r/3$, in particular, they are exactly solved   by
the standard solution with constant dilaton (see 
eqs. (\ref{c29}), (\ref{c30})), 
\beq
a \sim t^{1/2}, ~~~~~~~~~ 
\r=3p \sim a^{-4}, ~~~~~~~~~
\phi ={\rm const}, ~~~~~~~~~t \ra +\infty,
\label{214}
\eeq
describing decelerated expansion and decreasing curvature scale:
\beq
\dot a >0, ~~~~~~~~~ \ddot a <0,
~~~~~~~~~\dot H <0.
\label{215}
\eeq
This is exactly the  radiation-dominated solution of the standard
cosmological scenario, based on the Einstein equations. In string
cosmology, however, to this solution is associated a ``dual
complement", i.e. an additional solution which can be obtained
by applying on the background (\ref{214}) a 
time-reversal transformation $t \ra -t$, and the duality transformation
(\ref{211}): 
\beq
a \sim (- t)^{-1/2}, ~~~~~ \phi \sim -3 \ln (-t), ~~~~~
\r=-3p \sim a^{-2}, ~~~~~t\ra -\infty,
\label{216}
\eeq
This is still an exact solution of the equations (\ref{213}) (see the Appendix
C),  describing however accelerated (i.e. inflationary) expansion, with
growing dilaton and growing curvature  scale: 
\beq
\dot a >0, ~~~~~~~~~ \ddot a >0,
~~~~~~~~~\dot H>0.
\label{217}
\eeq
We note, for future reference, that accelerated expansion with
growing curvature is usually called ``superinflation" \cite{17}, or
``pole-inflation", to distinguish it from the more conventional
power-inflation, with decreasing curvature. 

The two solutions (\ref{214}) and 
(\ref{216}) provide a particular,  explicit  representation of the
scenario illustrated in Fig. 3, in the two asympotic regimes of
$t$ large and positive, and $t$ large and negative, respectively. 
The duality symmetry seems thus to provide an important motivation
for the pre-big bang scenario, as it leads naturally to introduce a
phase of growing curvature, and is a crucial ingredient for the
``bell-like" scenario of Fig. 3.  

It should be noted that pure scale factor duality, by itself, is not
enough to convert a phase of decreasing  into growing
curvature (see for instance Fig. 2, where it is clearly shown that $H$
and $\tilde H$, in the same temporal range, lead to the same evolution
of the curvature scale,  $H^2 \sim \tilde H^2$). Time reflection is thus 
necessarily  required, if we want to invert the curvature behaviour.
From this point of view, time-reversal symmetry is more important
than duality. 

In a thermodynamic context, however, duality by itself is able to
suggests the existence of a primordial cosmological phase with
``specular" properties with respect to the present, standard cosmological
phase \cite{18}. It must be stressed, in addition,  that it is typically in the
cosmology of extended objects that the phase of growing curvature may
describe accelerated expansion instead of contraction, and that the
growth of the curvature may be regularized,  instead of blowing up to a 
singularity. For instance, it is with the string dilaton \cite{8}, or with a
network of strings self-consistently coupled to the background
\cite{19}, that we are naturally lead to superinflation. Also, in 
quantum theories of extended objects, it is the minimal, fundamental
lenght scale of the theory that is expected to bound  the curvature,
and to drive  superinflation to a phase of constant, limiting curvature  
\cite{20} asymptotically approaching  de Sitter, as explicitly checked
in a string theory context \cite{21}.   Duality symmetries, on the
other hand,  are typical of extended objects (and of strings, in
particular), so that it is certainly justified to think of duality as of a
fundamental motivation and ingredient of the pre-big bang scenario. 

Duality is an important symmetry of modern theoretical
physics, and  to conclude this section I would like to present  an
analogy with another very important symmetry, namely
supersymmetry (see Table I). 

\begin{table}
\tabcolsep .07cm
\renewcommand{\arraystretch}{2.0}
\begin{center}
\begin{tabular}{|c||c||c|}
\hline
   &  {\bf SUPERSYMMETRY}  &  {\bf DUALITY + TIME-REVERSAL}   \\ \hline

 Pair  of partners      &    \{bosons, fermions\}   & 
\{growing curvature, decreasing curvature\}     \\ \hline

Known states   &  photons, gravitons, ...  &  decelerated, standard 
~post-big bang    \\ \hline

Predicted     &   photinos, gravitinos, ... & accelerated, inflationary  
~pre-big bang  \\ \hline 
\end{tabular}
\bigskip
\caption{Analogy between supersymmetry and duality.}
\end{center}
\end{table}

According to supersymmetry, to  any {\em bosonic state} is associated a
{\em fermionic partner}, and viceversa. From the existence of bosons that
we know to be present in Nature, if we believe in supersymmetry, we can 
predict the existence of fermions  not yet  observed, like the photino, the
gravitino, and so on. 

In the same way, according to duality and time-reversal, to any {\em
geometrical state} with decreasing curvature is associated a {\em dual
partner} with growing curvature. On the other hand, our Universe, at
present, is in the standard post-big bang phase, with decreasing
curvature. If we believe that duality has to be implemented, 
even approximately, in the course of the cosmological evolution, we can
then  predict the existence of a  phase, in the past, characterized by
growing curvature and by a typical pre-big bang evolution. 

\section{Kinematics: shrinking horizons}
\label{III}

If we accept, at least as a working hypothesis, the possibility that our
Universe had in the past a ``dual" complement,  with growing curvature,
we are lead to the second of the three questions listed in Section \ref{I}: 
is the kinematics of the pre-big bang phase still appropriate to solve the
problems of the standard inflationary scenario? The anwer is
positive, but in a non-trival way. 

Consider, for instance, the present cosmological phase. Today the dilaton
is expected to be constant, and the Universe should be appropriately
described by the Einstein equations.  The gravitational part of such 
equations contains two types of terms:  terms  controlling the geometric
curvature of a space-like section, evolving in time like  $a^{-2}$,  and
terms controlling the gravitational kinetic energy,  i.e.  the spacetime
curvature scale, evolving like $H^{2}$. According to present observations 
the  spatial curvature  term is non-dominant, i.e.    
\beq
r={a^{-2}\over H^2} \sim {\rm spatial~ curvature\over
spacetime ~curvature}
 ~\laq ~1.
\label{31}
\eeq

According to the standard cosmological solutions, on the other hand, the
above ratio must grow in time. In fact, by putting $a \sim t^\b$, 
\beq
r \sim \dot a ^{-2} \sim t^{2(1-\b)} ,
\label{32}
\eeq
so that $r$ keeps growing both in the matter-dominated ($\b=2/3$)
and in the radiation-dominated ($\b=1/2$) era. Thus, as we go back in
time,  $r$  becomes smaller and smaller, and when we set 
initial conditions (for instance, at the Planck scale) we have to impose an
enormous fine tuning of the spatial curvature term, with respect to
the other terms of the cosmological equations.
This is the so-called flatness problem. 

The problem can be solved if we introduce in the past 
a phase (usually called inflation), during which the value of $r$ was
decreasing, for a time long enough to compensate the subsequent growth
during the phase of standard evolution. It is important to
stress that  this requirement, in general, can be
implemented  by two physically different classes of
backgrounds. 

Consider for simplicity a power-law evolution of the scale
factor in cosmic time, with a power $\beta$, so that the time-dependence
of $r$ is the one given in eq. (\ref{32}). The two possible classes of
backgrounds corresponding to a decreasing $r$ are then the following:

\begin{itemize}

\item{}{\sl \underline{Class I}}: $a \sim t^\b, \b >1, t \ra +\infty$. This
class of backgrounds corresponds to what is conventionally called ``power
inflation", describing a phase of accelerated expansion and decreasing
curvature scale,  $\dot a >0, \ddot a >0, \dot H <0$. This class contains, as a
limiting case, the standard de Sitter inflation, $\b \ra \infty$, $a \sim
e^{Ht}$, $\dot H =0$, i.e. accelerated exponential expansion at constant
curvature. 
 
\item{}{\sl \underline{Class II}}: $a \sim (-t)^\b, \b <1, t \ra 0_-$. This
is the class of backgrounds corresponding to the string cosmology scenario.
There are two possible subclasess:

\begin{itemize}
\item[{\sl  \underline{IIa}}]: $\b<0$, describing superinflation,  i.e. 
accelerated expansion with growing curvature scale, 
$\dot a >0, \ddot a >0, \dot H >0$; 
\item[{\sl \underline{IIb}}]: $0<\b<1$, describing accelerated contraction 
and growing curvature scale, 
$\dot a <0, \ddot a <0, \dot H <0$. 
\end{itemize}
\end{itemize}

A phase of growing curvature, if accelerated like in the pre-big bang
scenario, can thus provide an unconventional, but acceptable, 
inflationary solution of the flatness problem (the same is true for the
other standard kinematical problems, see \cite{8}). It is important to
stress, in particular, that the two subclasses {\sl IIa},  {\sl IIb}, do not
correspond to different models, as they are simply different kinematical
representation of the {\em same} scenario in two {\em different frames},
the string frame (S-frame), in which the effective action takes the form 
(\ref{24}), 
\beq 
S(g,\phi)= - \int d^{d+1}x \sqrt{|g|}~
e^{-\phi}\left[R+ g^{\mu\nu}\pa_\mu\phi \pa_\nu \phi\right],
\label{33}
\eeq
and the Einstein frame (E-frame), in which the dilaton is minimally coupled
to the metric, and has a canonical  kinetic term: 
\beq 
S(\ti g,\ti \phi)= - \int d^{d+1}x \sqrt{|\ti g|}~
\left[\ti R-{1\over 2} \ti g^{\mu\nu}\pa_\mu\ti \phi \pa_\nu \ti
\phi\right] .
\label{34}
\eeq

In order to illustrate this point, we shall proceed in two steps. First we
will show that, through a field redefinition $g=g(\ti g,\ti \phi)$, $\phi=
\phi (\ti g,\ti \phi),$, it is always possible to move from the S-frame to
the E-frame; second, we will show that, by applying such a redefinition, a
superinflationary solution obtained in the S-frame becomes an accelerated
contraction in the E-frame, and viceversa. 

We shall consider, for simplicity, an isotropic, spatially flat background
with $d$ spatial dimensions, and we set:
\beq
g_{\mu\nu}= {\rm diag} \left( N^2, -a^2 \da_{ij}\right), ~~~~~~~
\phi=\phi(t), 
\label{35}
\eeq
where $g_{00}=N^2$ is to be fixed by an arbitrary choice of gauge. For this
background we get:
\bea
&&
\Ga_{0i}\,^j=H\da_i^j, ~~~~~~~ \Ga_{ij}\,^0= {a \dot a \over N^2} \da_{ij}, 
~~~~~~~ \Ga_{00}^0= {\dot N \over N}=F \nonumber\\
&&
R={1\over N^2} \left[2dFH -2d\dot H -d(d+1)H^2\right], 
\label{36}
\eea
and the S-frame action (\ref{33}) becomes
\beq
S(g,\phi)= - \int d^{d+1}x {a^d
e^{-\phi}\over N}\left[2dFH -2d\dot H -d(d+1)H^2 +\dot \phi^2\right]
\label{37} 
\eeq
Modulo a total derivative, we can eliminate the first two terms, and the
action takes the quadratic form
\beq
S(g,\phi)= - \int d^{d+1}x {a^d
e^{-\phi}\over N}\left[\dot \phi^2-2dH\dot\phi +d(d-1)H^2 \right]
\label{38} 
\eeq
where, as expected, $N$ plays the role of a Lagrange multiplier (no kinetic
term in the action).  

In the E-frame the variables are $\ti N, \ti a, \ti \phi$, and the
action(\ref{34}), after integration by part, takes the canonical form
\beq
S(\ti g, \ti \phi)= - \int d^{d+1}x {\ti a^d
\over \ti N}\left[-{1\over 2}\dot {\ti \phi}^2 +d(d-1)H^2\right] 
\label{39} 
\eeq
A quich comparison with eq. (\ref{38}) leads finally to the field
redefinition (no coordinate transformation!) connecting the Einstein and
String frame:
\beq
\ti a = a e^{-\phi/(d-1)}, ~~~~~~~~~
\ti N = N e^{-\phi/(d-1)}, ~~~~~~~~~
\ti \phi =\phi \sqrt{2\over d-1}.
\label {310}
\eeq
In fact, the above transformation gives
\beq
\ti H= H- {\dot \phi \over d-1}
\label{311}
\eeq
and, when inserted into eq. (\ref{39}), exactly reproduces the S-frame
action (\ref{38}). 

Consider now a superinflationary, pre-big bang solution obtained in the
S-frame, for instance the isotropic, $d$-dimensional vacuum solution
\beq
\fl a= (-t)^{-1/\sqrt{d}}, ~~~~~~~~~
e^\phi=  (-t)^{-(\sqrt{d}+1)}, ~~~~~~~~~
t<0, ~~~~~~~~~~t \ra 0_-
\label{312}
\eeq
(aee Appendix B, eqs. (\ref{b21}), (\ref{b22})), and look for the
corresponding E-frame solution. The above solution is valid in the
syncronous gauge, $N=1$, and if we choose, for instance, the
syncronous gauge also in the E-frame, we can fix $\ti N$ by the condition:
\beq
\ti N dt \equiv N e^{-\phi/(d-1)} dt = d \ti t,
\label{313}
\eeq
which defines the E-frame cosmic time, $\ti t$, as:
 \beq
d \ti t = e^{-\phi/(d-1)} dt .
\label{314}
\eeq
After integration
\beq
t \sim \ti t ^{d-1\over d +\sqrt{d}},
\label{315}
\eeq
and the transformed solution takes the form:
\beq
\fl \ti a= (-\ti t)^{1/{d}}, ~~~~~~~~~
e^{\ti \phi}=  (-\ti t)^{-\sqrt{2(d-1)\over d}}, ~~~~~~~~~
\ti t<0, ~~~~~~~~~ \ti t \ra 0_- .
\label{316}
\eeq
One can easily check that this solution describes accelerated contraction
with growing dilaton and growing curvature scale:
\beq
{d\ti a\over d \ti t}<0,~~~~~~~
{d^2\ti a\over d \ti t^2}<0,~~~~~~~
{d\ti H\over d \ti t}<0,~~~~~~~
{d\ti \phi\over d \ti t}>0. 
\label{317}
\eeq
The same result applies if we transform other isotropic solutions from the
String to the Einstein frame, for instance the perfect fluid solution of
Appendix C, eq. (\ref{c29}). We leave this simple exercise to the interested
reader. 

Having discussed the ``dynamical" equivalence (in spite of the kinematical
differences) of the two classes of string cosmology metrics, 
{\sl IIa} and {\sl IIb}, it seems appropriate at this point to stress the main
dynamical difference between standard inflation, {\sl
Class I} metrics,   and pre-big bang inflation, {\sl Class II} metrics. Such a
difference can be conveniently illustrated in terms of the 
proper size of the event horizon, relative to a given comoving observer. 

Consider in fact the the proper distance, $d_e(t)$, between the
surface of the event horizon and a comoving observer, at rest at the
origin of an isotropic, conformally flat background \cite{22}:
\beq
d_e(t)= a(t)\int _t^{t_M} dt' a^{-1} (t'),
\label{318}
\eeq
Here $t_M$ is the maximal allowed extension, towards the future, of the
cosmic time coordinate for the given background manifold. The above
integral coverges for all the above classes of accelerated (expanding or
contracting) scale factors. In the case of {\sl Class I} metrics we have, in
particular, 
\beq
d_e(t)= t^\beta\int _t^{\infty} dt' t'^{-\b}= {t\over \b-1} \sim H^{-1}(t)
\label{319}
\eeq
for power-law inflation ($\b >1, t>0$), and 
\beq
d_e(t)= e^{Ht}\int _t^{\infty} dt' e^{-Ht'}=  H^{-1}
\label{320}
\eeq
for de Sitter inflation. For {\sl Class II} metrics ($\b <1, t<0$) we have
instead 
\beq
d_e(t)= (-t)^\beta\int _t^{0} dt' (-t')^{-\b}= {(-t)\over 1-\b} \sim H^{-1}(t). 
\label{321}
\eeq
In all cases the proper size $d_e(t)$ evolves in time like the so-called
Hubble horizon (i.e. the inverse of the Hubble parameter), and then like
the inverse of the curvature scale. The size of the horizon is thus constant
or growing in standard inflation ({\sl Class I}), decreasing in pre-big bang
inflation ({\sl Class II}), both in the S-frame and in the E-frame. 

\begin{figure}
\centerline{\epsfxsize=8.0cm
\epsffile{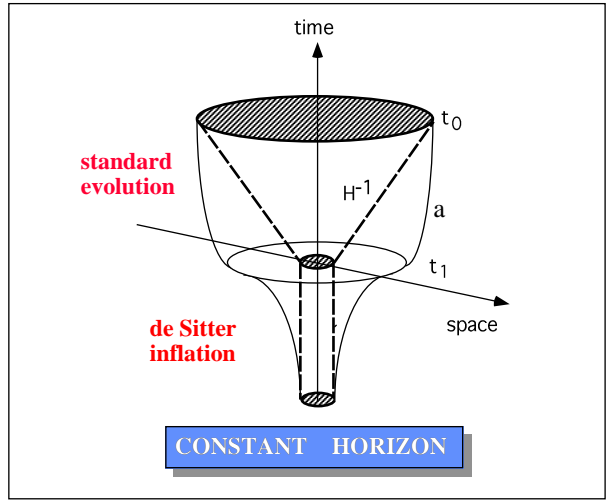}}
\caption{\sl Qualitative evolution of the Hubble horizon (dashed curve)
and of the scale factor (full curve) in the standard inflationary scenario.} 
\label{fig4}
\end{figure}

Such an important difference is clearly illustrated in Fig. 4 and Fig. 5,
where the dashed lines represent the evolution of the horizon, the full
lines the evolution of the scale factor. The shaded area at time $t_0$
represents the portion of Universe inside our present Hubble radius. As
we go back in time, according to the standard scenario, the horizon
shrinks linearly, ($H^{-1} \sim t$), but the decrease of the scale factor is
slower so that, at the beginning of the phase of standard evolution
($t=t_1$), we end up with a causal horizon much smaller than the  portion
of Universe that we presently observe. This is the well known ``horizon
problem" of the standard scenario. 

In Fig. 4 the phase of standard evolution is preceeded in time by a phase
of standard de Sitter inflation. Going back in time, for $t<t_1$, the scale
factor keeps shrinking, and our portion of  Universe ``re-enters" inside
the Hubble radius during a phase of constant (or slightly growing in time)
horizon. 

In Fig. 5 the standard evolution is preceeded in time by a phase
of pre-big bang inflation, with growing curvature . The  Universe
``re-enters"  the Hubble radius during a phase of shrinking  horizon. To
emphasize the difference, I have plotted the evolution of the scale factor
both in the expanding S-frame, $a(t)$, and in the contracting E-frame, $\ti
a (t)$. Unlike in standard inflation, the proper size of the initial portion of
the Universe may be very large in strings (or Planck) units, {\em but not
larger than the initial horizon} itself \cite{23}, as emphasized in the
picture. The initial horizon is large because the initial curvature scale is
small, in string units, $H_i \ll 1/\la_s$. 

\begin{figure}
\centerline{\epsfxsize=8.0cm
\epsffile{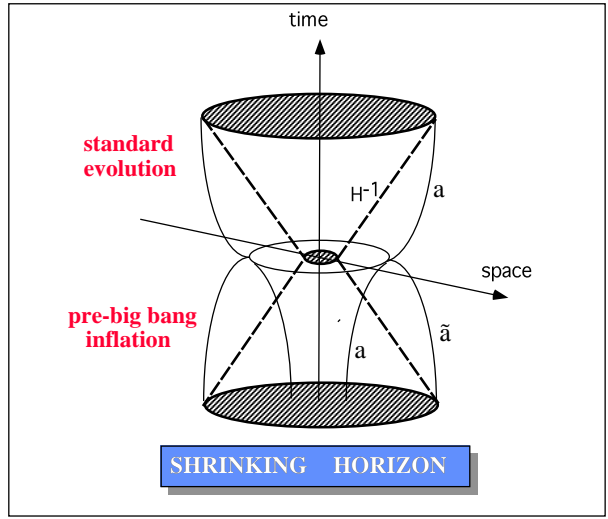}}
\caption{\sl Qualitative evolution of the Hubble horizon (dashed curve)
and of the scale factor (full curve) in the pre-big bang inflationary
scenario, in the S-frame, $a(t)$, and in the E-frame, $\ti a(t)$.} 
\label{fig5}
\end{figure}

This is a basic consequence of the choice of the initial state which, in the
pre-big bang scenario, approaches the flat, cold and empty string
perturbative vacuum \cite{8}, and which is to be contrasted to the
extremely curved, hot and dense initial state of the standard scenario,
characterizing a Universe which starts inflating at the Planck scale, 
$H_i \sim 1/\la_p$.

\section{Open problems and phenomenological consequences}
\label{IV}

In order to give a honest presentation of the pre-big bang scenario, it
is fair to say that the string cosmology models are not free from various
(more or less important) difficulties, and that many aspects of the
scenario are still unclear. A detailed discussion of such aspects is outside
the purpose of this paper, but I would like to mention here at least three
important open problems. Presented in ``time-ordered" form (from the
beginning to the end of inflation) they are the following.

\begin{itemize}

\item{}The first concerns the initial conditions, and in particular 
the decay of the string perturbative vacuum. The question is 
whether or not  the ``switching on" of a long inflationary phase requires
fine-tuning. Originally raised in \cite{24}, this problem was recently
re-proposed as a fundamental difficulty of the pre-big bang scenario
\cite{25} (see however \cite{23,26,27,28}). 

\item{}The second concerns the transition from the pre- to the post-big
bang phase, which is expected to occur in the high curvature and strong
coupling regime. There is a quantum cosmology approach, based on the
scattering of the Wheeler-De Witt wave function in minisuperspace
\cite{7}, but the problem seems  to require, in general, the introduction of
higher derivative ($\ap$) and quantum loop  corrections \cite{21,29} in
the string effective action (see however \cite{15,16}).

\item{}The third problem concerns the final matching to the standard
Friedman-Robertson-Walker phase, with a transition from the
dilaton-dominated to the radiation-dominated regime, and all the
associated problems of dilaton oscillations, re-heating, pre-heating,
particle production, entropy production \cite{30}, and so on.
\end{itemize}

All these problems are under active investigation, and 
further work is certainly needed for a final answer. However, even
assuming that all the problems will be solved in a satisfactory way, we are
left eventually with a further question, the third one listed in the
Introduction, which is the basic question (in my opinion). 
Are there phenomenological consequences that can discriminate string
cosmology from the other inflationary scenarios? and, in particular, are
such consequences observable (at least in principle) ? 

The answer is positive. There are many phenomenological differences,
even if all the differences seem to have the same ``common denominator",
i.e. the fact that the quantum fluctuations of the background fields are
amplified in different models with different spectra. The spectrum, in
particular, tends to follow the behaviour of the curvature scale during the
phase of inflation. In the standard scenario the curvature is constant or
decreasing, so that the spectrum tends to be flat, or decreasing with
frequency. In string cosmology the curvature is growing, and the
spectrum tends to grow with frequency. 

In the following sections I will discuss in detail this effect for the case of
tensor metric perturbations. Here I would like to note  that the
phenomenological consequences of the pre-big bang scenario can be
classified into three different types, depending on the possibility of their 
observation: {\sl Type I} effects, referring to observations to be
performed in a not so far future (20-30 years?); {\sl Type II} effects,
referring to observations to be performed in a near future (a few years); 
{\sl Type III} effects, referring to observations already (in part)
performed. To conclude this very quick presentation of the pre-big bang
scenario, let me give one example for each type of phenomenological
effect. 

\begin{itemize}

\item{}{\sl \underline{Type I}}: the production of a relic
graviton background that, in the frequency range of conventional
detectors ($\sim 10^2 - 10^3$ Hz), is much higher (by 8 -- 9 orders of
magnitude) than the background expected in conventional inflation
\cite{8,31,32,33}. The sensitivity of the presently operating gravitational
antennas is not enough to detect it, however, and we have to wait for the
advanced, second generation interferometric detectors (LIGO \cite{34},
VIRGO \cite{35}), or for interferometers in space (LISA \cite{36}). 

\item{}{\sl \underline{Type II}}: the large scale CMB anisotropy ``seeded"
by the inhomogeneous fluctuations of a massless \cite{37} or massive
\cite{38} axion background. Metric fluctuations are indeed too small, on the
horizon scale, to be responsible for the temperature anisotropies
detected by COBE \cite{39}; the axion spectrum, on the contrary, can be
sufficiently flat \cite{40} for that purpose. Such a different origin of the
anisotropy may lead to non-gaussianity, or to differences (with respect to
the standard inflationary scenario) in the height and position of the first
Doppler peak of the spectrum \cite{41}. Such differencs could be soon
confirmed, or disproved, by the planned satellite observations (MAP
\cite{42}, PLANCK \cite{43}, ... ). 

\item{}{\sl \underline{Type III}}: the production of primordial magnetic
fields strong enough to ``seed" the galactic dynamo, and to explain the
origin of the cosmic magnetic fields observed on a large (galactic,
intergalactic) scale \cite{44}. In the standard inflationary scenario, in
fact, the amplification of the vacuum fluctuations of the electromagnetic
field is not efficient enough \cite{45}, because of the conformal invariance
of the Maxwell equations. In string cosmology, on the contrary,
the electromagnetic field is also coupled to the dilaton, and the
fluctuations are amplified by the accelerated growth of the dilaton during
the phase of pre-big bang evolution. 
\end{itemize}

Finally, I wish to mention a further important phenomenogical effect,
typical of string cosmology (and that I do not know how to classify within
the tree types defined above, however): dilaton production, i.e. the
amplification of the dilatonic fluctuations of the vacuum, and the
formation of a cosmic background of relic dilatons \cite{46}. 

The possibility of detecting such a background is strongly dependent on the
value of the dilaton mass, that we do not known, at present. If dilatons
are massless \cite{47}, then the amplitude and the spectrum of the relic
background should be very similar to those of the graviton background,
and the relic dilatons could be possibly detected, in the future, by
gravitational antennas able to respond to scalar modes, unless their
coupling to bulk matter is too small \cite{47}, of course. 

If dilatons are massive, the mass has to be large enough to be compatible
with existing tests of the equivalence principle and of macroscopic
gravitational forces. In addition, there is a rich phenomenology of
cosmological bounds, which leaves open only two possible mass windows
\cite{46}. Interestingly enough, however, in the allowed light mass sector
the dilaton lifetime is longer than the present age of the universe, and the
dilaton fraction of critical energy density ranges from $0.01$ to $1$: 
in this context, the dilaton becomes  a new, interesting dark matter
candidate (see \cite{48} for a detailed discussion of the allowed mass
windows, and of the possibility that light but non-relativistic dilatons
could represent today a significant fraction of dark matter on a
cosmological scale).  I have no idea, however, of how to detect directly
such a massive dilaton  background, because the mass is light, but it is
heavy enough ($\gaq~10^{-4}$eV) to be far outside the sensitivity range of
resonant gravitational detectors. 

The rest of this lecture will be devoted to discuss various theoretical and
phenomenological aspects of graviton production,  in a
general cosmological context  and, in particular, in the context of the
pre-big bang scenario. Let me start by recalling some basic
notions of cosmological perturbation theory, which are required for the
computation of the graviton spectrum. 

\section{Cosmological perturbation theory}
\label{V}

The standard approach to cosmological perturbation theory is to start
with a set of non-perturbed equations, for instance the Einstein or the
string cosmology equations,
\beq
G_{\mu\nu}= T_{\mu\nu}, 
\label{51}
\eeq
to expand the meric and the matter fields around a given background
solution, 
\beq
g_{\mu\nu} \ra g_{\mu\nu}^{(0)}+ \da^{(1)} g_{\mu\nu},
~~~~~~T_{\mu\nu} \ra T_{\mu\nu}^{(0)}+ \da^{(1)} T_{\mu\nu}, 
~~~~~~G_{\mu\nu}^{(0)}= T_{\mu\nu}^{(0)}, 
\label{52}
\eeq
and to obtain, to first order, a linearized set of equations describing the
classical evolution of perturbations,
\beq 
\da^{(1)}G_{\mu\nu}= \da^{(1)} T_{\mu\nu}.
\label{53}
\eeq
In principle, the procedure is simple and straightforward. In practice,
however, we have to go through a series of formal steps, that I list here
in ``chronological" order: 

\begin{itemize}

\item{}choice of the ``frame";

\item{}choice of the ``gauge";

\item{}normalization of the amplitude;

\item{}computation of the spectrum.

\end{itemize}

\subsection{Choice of the frame}
\label{Va}

The choiche of the frame is the choice of the basic set of fields (metric
included) used to parametrize the action. The action, in general, can be
expressed in terms of different fields. In string cosmology, for instance,
there is a preferred frame, the S-frame, in which the lowest order
gravi-dilaton action takes the form (\ref{33}). It is preferred because the
metric appearing in the action is the same as the sigma-model metric to
which test strings are minimally coupled (see Appendix A): with respect to
this metric, the motion of free strings is then  geodesics. It is always
possible, however, through the field redefinition
\beq
\ti g_{\mu\nu}= g_{\mu\nu} e^{-2\phi/(d-1)}, ~~~~~~~~~~~
\ti \phi= \sqrt{2\over d-1} \phi , 
\label{54}
\eeq
to introduce the more conventional E-frame (\ref{34}) in which the dilaton
is minimally coupled to the metric, with a canonical kinetic term. 

In the two frames the field equations are different, and the perturbation
equations are also different. This seems to rise a potential problem: which
frame is to be used to evaluate the physical effects of the
cosmological perturbations?

The problem is only apparent, however, because physical observables (like
the perturbation spectrum) are the same in both frames. The reason is
that there is a compensation between the different perturbation
equations and the different background solution around which we expand.
A general proof of this result can be given by using the notion of
canonical variable (see the Subsection \ref{Vc}). Here I will give only an
explicit example for tensor perturbations in a $d=3$,  isotropic
and spatially flat background.

Let me start in the E-frame, with the background equations:
\beq
R_{\mu\nu}= {1\over 2} \pa_\mu \phi \pa_\nu \phi, 
\label{55}
\eeq
referring to the ``tilded" variables (\ref{54}) (I will omit  the ``tilde", for
simplicity,  and I will explicitly reinsert it at the end of the computation). 
Consider the transverse, traceless part of metric perturbations:
\beq
\fl 
\da^{(1)}\phi=0, ~~~~
\da^{(1)} g_{\mu\nu} = h_{\mu\nu}, ~~~~~
\da^{(1)} g^{\mu\nu} = -h^{\mu\nu}, ~~~~~
\nabla_\nu h_\mu\,^\nu =0=h_\mu\,^\nu
\label{56}
\eeq  
($\nabla_\mu$ denotes covariant differentiation with respect to the
unperturbed metric $g$, and the indices of $h$ are also raised and
lowered with $g$). The perturbation of the background equations
gives:
\beq
\da^{(1)} R_\mu\,^\nu =0.
\label{57}
\eeq
We can work in the syncronous gauge, where
\bea
&& 
g_{00}=1, ~~~~~g_{0i}=0, ~~~~~ g_{ij}=-a^2\da_{ij}, \nonumber \\
&&
h_{00}=0, ~~~~~h_{0i}=0, ~~~~~ g^{ij}h_{ij}=0, ~~~~~ \pa_j h_i\,^j =0 .
\label{58}
\eea
To first order in $h$ we get
\bea
&&
\da^{(1)} \Ga_{0i}\,^j={1\over 2}\hp_i\,^{j}, ~~~~ 
\da^{(1)}\Ga_{ij}\,^0=-{1\over 2}\hp_{ij}, \nonumber \\ 
&&
\da^{(1)}\Ga_{ij}\,^k={1\over 2}\left(\pa_i h_j\,^k+
\pa_j h_i\,^k-\pa^k h_{ij}\right). 
\label{59}
\eea
The $(0,0)$ component of eq. (\ref{57}) is trivially satisfied (as well as the
perturbation of the scalar field equation); the $(i,j)$ components, by using
the identities (see for instance \cite{53})
\bea
&&
g^{jk}\hp_{ik}=\hp_i\,^j +2 H h_i\,^j , \nonumber\\
&&
g^{jk}\hpp_{ik}=\hpp_i\,^j +2 \dot H h_i\,^j +4H\hp_i\,^j +4H^2h_i\,^j, 
\label{510}
\eea
give
\beq
\da^{(1)}R_{i}\,^j=-{1\over 2}\left(\hpp_i\,^j
+3H \hp_i\,^j -{\nabla^2\over a^2} h_i\,^j\right)\equiv 
-{1\over 2} \square h_i\,^j  =0.
\label{511}
\eeq
In terms of the conformal time coordinate, $d\eta = dt/a$, this 
wave equation  can be finally rewritten, for each polarization mode, as
\beq
\ti h^\se +2 {\ti a' \over \ti a} \ti h' - \nabla^2 \ti h =0 ,
\label{512}
\eeq
(where I have explicitly re-inserted the tilde, and where a prime denotes
differentiation with respect to the conformal time, which is the same in
the Einstein and in the String frame, according to eqs. (\ref{310}),
(\ref{314})). 

Let us now repeat the computation in the S-frame, where the background
equations for the metric (eq. (\ref{c10}) with no
contribution from $H_{\mu\nu\a}, T_{\mu\nu}, V$ and $\sg$) can be
written explicitly as 
\beq
R_\mu\,^\nu + g^{\nu\a}\left(\pa_\mu \pa_\a \phi - \Ga _{\mu\a}\, ^\rho
\pa_\rho \phi\right)=0.
\label{513}
\eeq
Perturbing to first order, 
\beq
\da^{(1)}R_\mu\,^\nu -\left(\da^{(1)} g^{\nu\a}\Ga_{\mu\a}\, ^0 +
g^{\nu\a}\da^{(1)} \Ga_{\mu\a}\, ^0 \right) \dot \phi=0. 
\label{514}
\eeq
The $(0,0)$ component, as well as the perturbation of the dilaton equation,
are trivially satisfied. The $(i,j)$ components, using again the identities
(\ref{510}), lead to \cite{31}:
\beq
\square h_i\,^j -\dot \phi \dot h_i\,^j =0.
\label{515}
\eeq
In conformal time, and for each polarization component,
\beq
h^\se +\left(2 { a' \over  a}- \phi' \right)  h' - \nabla^2  h =0 
\label{516}
\eeq

This last equation seems to be different frm the E-frame equation
(\ref{512}). Recalling, however,  the relation (\ref{310}) between $a$ and
$\ti a$, we have
\beq
2 {\ti a' \over \ti a}= 2 { a' \over  a}- \phi' ,
\label {517}
\eeq
so that we have the same equation for $h$ and
$\ti h$, the same solution, and  the same spectrum when the solution
is expanded in Fourier modes. 
The perturbation analysis is thus {\em frame-independent}, and we can
safely choose the more convenient frame to compute the spectrum. 

\subsection{Choice of the gauge}
\label{Vb}

 The second step is the choice of the gauge, i.e. the choice of the
coordinate system within a given frame. The perturbation spectrum is of
course gauge-independent, but the the perturbative analysis {\em is not},
in general. It is possible, in fact, that the validity of the linear
approximation is broken in a given gauge, but still valid in a different,
more appropriate gauge. 

Since this effect is particularly important, let me give, in short, an explicit
example for the scalar perturbations of the metric tensor in a $d=3$, 
isotropic and conformally flat background, in the E-frame (I will omit the
tilde, for simplicity). The perturbed metric, in the so-called longitudinal
gauge, depends on the two Bardeen potentials $\Phi$ and $\Psi$ as
\cite{49}:
\beq
ds^2= 
a^2\left[\left(1+2\Phi\right) d\eta^2 -
\left(1-2\Psi\right)dx_i^2\right]. 
\label{518}
\eeq
By perturbing the Einstein equations (\ref{55}), the dilaton equation, and
combining the results for the various components, one obtains to first
order that $\Phi=\Psi$, and that the metric fluctuations satisfy the 
equation: 
\beq
\Phi^\se +6{a'\over a}\Phi'- \nabla^2\Phi=0.
\label{519}
\eeq
We now consider the particular, exact solution of the vacuum string
cosmology equations in the E-frame, 
\beq
a(\eta)= |\eta|^{1/2}, ~~~~~~~~~~~
\phi(\eta)= -\sqrt{3} \ln |\eta|, ~~~~~~~~~~~\eta \ra 0_-,
\label{520}
\eeq
corresponding to a phase of accelerated contraction and
growing dilaton (i.e. the pre-big bang  solution (\ref{316}), written in
conformal time, for $d=3$). For this background, the perturbation eq.
(\ref{519}) becomes a Bessel equation for the Fourier modes $\Phi_k$, 
\beq
\Phi_k^\se +{3\over \eta}\Phi_k'+k^2\Phi_k=0,
~~~~~~~~~~~\nabla^2\Phi_k= -k^2 \Phi_k,
\label{521}
\eeq
and the asymptotic solution, for modes well outside the horizon
($|k\eta|\ll 1$), 
\beq
\varphi_k = A_k \ln |k\eta| + B_k |k\eta|^{-2}
\label{522}
\eeq
contains a growing part which blows up ($\sim \eta^{-2}$) as the
background approaches the high curvature regime ($\eta \ra 0_-$). In this
limit the linear approximation breaks down, so that the longitudinal
gauge is not in general consistent with the perturbative expansion around
a homogeneous, inflationary pre-big bang background, as scalar
inhomogeneities may become  too large. 

In the same background (\ref{520}) the problem is absent, however, for
tensor perturbations, since their growth outside the horizon is only
logarithmic. From eq. (\ref{512}) we have in fact the asymptotic solution
\beq
h_k = A_k+B_k\ln |k\eta| , ~~~~~~~ |k\eta| \ll 1 .
\label{523}
\eeq
This may suggests that the breakdown of the linear approximation, for
scalar perturbations, is an artefact of the longitudinal gauge. This is
indeed confirmed by the fact that, in a more appropriate off-diagonal (also
called ``uniform curvature" \cite{50}) gauge, 
\beq
ds^2= 
a^2\left[\left(1+2\varphi\right) d\eta^2 -
dx_i^2- 2\pa_iB dx^i d\eta \right], 
\label{524}
\eeq
the growing mode is ``gauged down", i.e. it is suppressed enough to
restore the validity of the linear approximation \cite{51} (the off-diagonal
part of the metric fluctuations remains growing, but the growth is
suppressed in such a way that the amplitude, normalized to the vacuum
fluctuations, keeps smaller than one for all scales $k$, provided the
curvature is smaller than one in string units). This result is also
confirmed by a covariant and gauge invariant computation of the
spectrum, according to the formalism developed by Bruni and Ellis
\cite{51a}. 

It should be stressed, however,  that the presence of a growing mode, and
the need for choosing an appropriate gauge, is a problem typical of the
pre-big bang scenario. In fact, let us come back to tensor perturbations,
in the E-frame: for a generic accelerated background the scale factor can
be parametrized in conformal time with a power $\a$, as follows:
\beq
a= (-\eta)^\a, ~~~~~~~~~~~~  \eta \ra 0_- ,
\label{525}
\eeq
and the perturbation equation (\ref{512}) gives, for each Fourier mode,
the Bessel equation
\beq
h_k^\se +{2 \a\over \eta}h_k'+k^2h_k=0,
\label{526}
\eeq
with asymptotic solution, outside the horizon ($|k\eta| \ll 1$):
\beq
h_k =A_k+B_k\int^\eta {d\eta'\over a^2(\eta')} = A_k
+B_k\left|\eta\right|^{1-2\a}.
\label{527}
\eeq
The solution tends to be constant for $\a<1/2$, while it tends to
grow for $\a>1/2$. It is now an easy exercise to re-express the scal
factor (\ref{525}) in cosmic time,
\beq
dt=a d\eta, ~~~~~~~~~~~~~~
a(t) \sim |t|^{\a/(1+\a)},
\label{528}
\eeq
and to chech that, by varying $\a$, we can parametrize all types of
accelerated backgrounds introduced in Section \ref{III}: accelerated
expansion (with decreasing, constant and growing curvature), and
accelerated contraction, with growing curvature (see Table II). 

\begin{table}
\tabcolsep .4cm
\renewcommand{\arraystretch}{2.0}
\begin{center}
\begin{tabular}{|c||c||c|}
\hline
$\a<-1$   & power-inflation  &  $\dot a >0, \ddot a >0, \dot H<0 $  \\ 
\hline
$\a=-1 $   &   de Sitter    & $\dot a >0, \ddot a >0, \dot H=0  $  \\ 
\hline
$-1<\a<0  $ &  \rm super-inflation  & $\dot a >0, \ddot a >0, \dot H >0  $\\ 
\hline
$\a>0 $    &   accelerated~contraction & $\dot a <0, \ddot a <0, \dot H<0 $\\ 
\hline 
\end{tabular}
\bigskip
\caption{The four classes of accelerated backgrounds.}
\end{center}
\end{table}

In the standard, inflationary scenario the metric is expanding, $\a <0$, so
that the amplitude $h_k$ is frozen outside the horizon. In the pre-big bang
scenario, on the contrary, the metric is contracting in the E-frame, so that
$h_k$ may grow if the contraction is fast enough, i.e. $\a >1/2$ (in fact,
the growing mode problem  was first pointed out in the context of
Kaluza-Klein inflation and dynamical dimensional reduction \cite{52},
where the internal dimensions are contracting). For the low-energy string
cosmology background (\ref{520}) we have $\a=1/2$, the growth is simply
logarithmic (see (\ref{523})), and the linear approximation can be applied
consistently,  provided the curvature remains bounded by the string scale
\cite{51}. But for $\a >1/2$ the growth of the  amplitude may require a
different gauge for a consistent linearized description. 

\subsection{Normalization of the amplitude}
\label{Vc}

The linearized equations describing the classical evolution of
perturbations can be obtained in two ways:

\begin{itemize}
\item{}by perturbing directly the background equations of motion;
\item{}by perturbing the metric and the matter fields to first order, by
expanding the action up to terms quadratic in the first order fluctuations, 
\beq
g \ra g +\da^{(1)} g, ~~~~~~~~~~~~
\da^{(2)} S \equiv S\left[ \left( \da^{(1)} g\right)^2\right] ,
\label {529}
\eeq
and then by varying the action with respect to the fluctuations. 
\end{itemize}

\noindent
The advantage of the second method is to define  the
so-called ``normal modes" for the oscillation of the system \{gravity +
matter sources\}, namely the variables which diagonalize the kinetic terms
in the perturbed action, and satisfy canonical commutation relations when
the fluctuations are quantized. Such canonical variables are required, in
particular, to normalize perturbations to a spectrum of quantum,
zero-point fluctuations, and to study their amplification from the vacuum
state up to the present state of the Universe. 

Let us apply such a procedure to tensor perturbations, in the S-frame, for
a $d=3$ isotropic background. In the syncronous gauge, the transverse,
traceless, first order metric perturbations  $h_{\mu\nu}=\da^{(1)}
g_{\mu\nu}$ satisfy eq. (\ref{58}). We expand all terms of the low energy
gravi-dilaton action (\ref{33}) up to order $h^2$:
\bea
&&
\da^{(1)} g^{\mu\nu}=-h^{\mu\nu}, ~~~~~~~~~~~~ \da^{(2)}
g^{\mu\nu}=h^{\mu\a}h_\a\,^\nu , \nonumber \\
&&
\da^{(1)} \sqrt{-g}=0, ~~~~~~ \da^{(2)} \sqrt{-g}=-{1\over
4}\sqrt{-g}h_{\mu\nu}h^{\mu\nu},  
\label{530}
\eea
and so on for $\da^{(1)}R_{\mu\nu}$,  $\da^{(2)}R_{\mu\nu}$ (see for
instance \cite{53}). By using the background equations, and integrating by
part, we finally arrive at the quadratic action 
\beq
\da^{(2)} S=
{1\over 4}\int d^4xa^3 e^{-\phi} \Bigg(\dot h_i^j\dot h_j^i + 
h_i^j{\nabla\over a^2} h_j^i \Bigg).
\label{531}
\eeq
By separating the two physical polarization modes, i.e. the standard
``cross" and ``plus" gravity wave components,
\beq
 h_i^j h_j^i = 2 \left( h_+^2 + h_\times^2\right),
\label{532}
\eeq
we get, for each mode (now generically denoted with $h$), the effective
scalar action
\beq
\da^{(2)} S=
{1\over 2}\int d^4xa^3 e^{-\phi} \left(\dot h^2+ 
h{\nabla\over a^2} h \right),
\label{533}
\eeq
which can be rewritten, using conformal time, as
\beq
\da^{(2)} S=
{1\over 2}\int d^3xd\eta a^2 e^{-\phi} \left( h^{\prime 2}+ 
h{\nabla} h \right).
\label{534}
\eeq
The variation with respect to $h$ gives finally eq. (\ref{516}), i.e.
the same  equation  obtained by perturbing directly the background
equations in the S-frame. 

The above action describes a scalar field $h$, non-minimally coupled to a
time-dependent external field, $a^2 e^{-\phi}$ (also called ``pump field").
In order to impose the correct quantum normalization to vacuum
fluctuations, we introduce now the so-called ``canonical variable" $\psi$,
defined in terms of the pump field as 
\beq
\psi= z h , ~~~~~~~~~~~~~~~
z =ae^{-\phi/2}. 
\label{535}
\eeq
With such a definition the kinetic term for $\psi$ appears in the standard
canonical form: for each mode $k$, in fact, we get the action
\beq
\da^{(2)} S_k=
{1\over 2}\int d\eta  \left( \psi_k^{\prime 2}- 
k^2\psi_k^2+ {z^\se\over z}\psi_k^2 \right),
\label{536}
\eeq
and the corresponding canonical evolution equation: 
\beq
\psi_k^\se +\left[k^2-V(\eta)\right]\psi_k =0, ~~~~~~~~~~~
V(\eta)={z^\se\over z},
\label{537}
\eeq
which has the form of a Schrodinger-like equation, with an effective
potential depending on the external pump field. This form of the canonical
equation, by the way, is the same for all types of perturbations (with
different potentials, of course). What is important, in our context, is that
for an accelerated inflationary background $V(z) \ra 0$ as $\eta \ra
-\infty$. This means that, asymptotically, the canonical variable satisfies
the free-field oscillating equation
\beq
\eta \ra -\infty, ~~~~~~~~~~~
 \psi_k^\se +k^2\psi_k =0,
\label{538}
\eeq
and can be normalized to an initial vacuum fluctuation spectrum, 
\beq
\eta \ra -\infty,  ~~~~~~~~~~~
\psi_k={1\over \sqrt{2k}} e^{- ik\eta},
\label{539}
\eeq
in such a way as to satisfy the free field canonical commutation relations, 
$[\psi_k, \psi^{*'}_j]=i\da_{kj}$. The normalization of $\psi_k$ then fixes 
the normalization of the metric  variable $h_k=\psi_k/z$. 

It is important to stress that there is no need to introduce the canonical
variable to study the classical evolution of perturbations, but that such
variable is needed for the initial normalization to a vacuum fluctuation
spectrum. We can also normalize perturbation in a different way, of
course but in that case we are studying the amplification not of the
vacuum fluctuations, but of a different spectrum \cite{53a}. 

At this point, two remarks are in order. The first concerns the
frame-independence of the spectrum. The above procedure can also be
applied in the E-frame, to define a canonical variable $\ti \psi$: one then
obatins for $\ti \psi_k$ the canonical equation (\ref{537}), with a pump
field that depends only on the metric, $\ti z =\ti a$. However, by using the
conformal transformation connecting the two frames, it turns out that the
two pump fields  are the same, $\ti z =\ti a= a e^{-\phi/2}=z$, so that 
for $\psi$ and $\ti \psi$  we have the same potential, the same evolution
equation, the same solution, and thus the same spectrum. 

The second remark is that the canonical procedure can be applied to any
action, and in particular to the string effective action including higher
curvature  corrections of order $\ap$, which can be written as \cite{21}: 
\beq
S=\int d^{4}x \sqrt{-g}e^{-\phi} 
\left\{- R-
\pa_\mu \phi\pa^\mu \phi+{\ap \over 4} \left[R^2_{GB} - 
(\pa_\mu\phi\pa^\mu\phi)^2\right]\right\}
\label{540}
\eeq
where  $R^2_{GB} 
\equiv R_{\mu\nu\a\b}^2-4  R_{\mu\nu}^2+
R^2$ is the Gauss-Bonnet invariant 
(we have chosen a convenient field redefinition that removes terms
with  higher-than-second derivatives from the equations of
motion, see Appendix A). From the quadratic perturbed action we obtain
$\ap$ corrections to the pump fields. The canonical equation turns out to
be the same as before, but with a $k$-dependent effective potential
\cite{53}, and such an equation can be used to estimate the effects of
the higher curvature corrections on the amplification of tensor
perturbations.  A numerical integration \cite{53}, in which the metric 
fluctuations are expanded around the high-curvature background solution
of ref. \cite{21}, leads in particular to the results illustrated in Fig. 6. 

\begin{figure}
\centerline{\epsfxsize=8.0cm
\epsffile{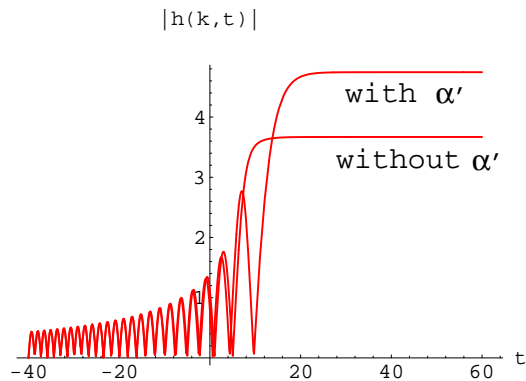}}
\caption{Amplification of tensor fluctuations, 
with and without the higher-curvature  corrections included in the
canonical perturbation equation.} 
\label{fig6}
\end{figure}

The qualitative behaviour is similar, both with and without $\ap$
corrections in the perturbed equations: the fluctuations are oscillating
inside the horizon and frozen outside the horizon, as usual. However, the
final amplitude is enhanced when $\ap$ corrections are included, and this
suggests that the energy  spectrum of the gravitational radiation,
computed with the low-energy perturbation equation, may represent a
sort of lower bound on the total amount of produced gravitons. 

\subsection{Computation of the spectrum}
\label{Vd}

The final, amplified perturbation spectrum is to be obtained from
the solutions of the canonical equation (\ref{537}). In order to solve such
an equation we need explicitly the effective potential $V[z(\eta)]$ which,
in general,   vanishes asymptotically at large postive and negative values
of the conformal time. Consider, for instance, the tensor perturbation
equation in the E-frame, so that the pump field is simply the scale factor.
The typical cosmological bckground in which we are interested in should 
describe a transition fom an initial accelerated, inflationary evolution, 
\beq
\eta \ra -\infty, ~~~~~~~~~~
a \sim |\eta|^\a, ~~~~~~~~~~
V \sim \eta^{-2},
\label{541}
\eeq
to a final standard, radiation-dominated phase,
\beq
\eta \ra +\infty, ~~~~~~~~~~
a \sim \eta, ~~~~~~~~~~
V =0.
\label{542}
\eeq
In this context, the evolution of fluctuations, initially normalized as in eq.
(\ref{539}), can be described as a scattering of the canonical variable by an
effective potential, according to the Schrodinger-like perturbation
equation (\ref{537}) (see Fig. 7). 

However, the differential variable in eq. (\ref{537}) is (conformal) time,
not space. As a consequence, the eigenfrequencies represent (comoving)
energies, not momenta. Thus, even normalizing the initial state to a
positive frequency mode, as in eq. (\ref{539}), the final state is in general a
mixture of positive and negative frequency modes, i.e. of positive and
negative energy states, 
\beq
\eta \ra +\infty , ~~~~~~~~~
\psi_{\rm out} \sim c_+e^{-ik\eta} + c_-e^{+ik\eta}.
\label{543}
\eeq

\begin{figure}
\centerline{\epsfxsize=8.0cm
\epsffile{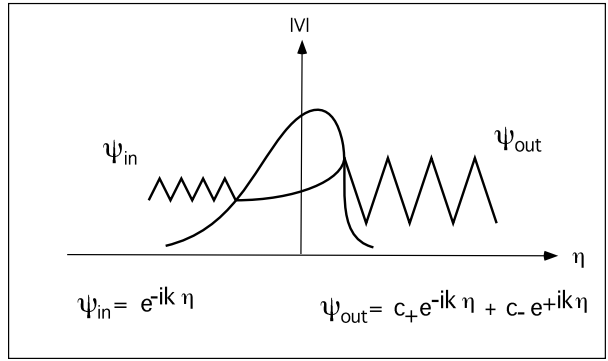}}
\caption{Scattering and amplification of the canonical variable.} 
\label{fig7}
\end{figure}

In a quantum field theory context, such a mixing represents a process of
pair production from the vacuum. The coefficients $c_\pm$ are the
so-called Bogoliubov coefficients, parametrizing a 
unitary  transformation between $|{\rm
in}\rangle$ and $|{\rm out}\rangle$ states. In matrix form, they connect 
the set of $|{\rm in}\rangle$ annihilation and creation
operators, $\{\psi_{\rm in}, b_k, b_k^\dagger\}$ to the out ones 
$\{\psi_{\rm out}, a_k, a_k^\dagger\}$, as follows:
\beq
a_k=c_+b_k+c_-^\ast b_{-k}^\dagger, ~~~~~~
a_{-k}^\dagger=c_-b_k+c_+^\ast b_{-k}^\dagger .
\label{544}
\eeq
Thus, even starting from the vacuum, 
\beq
\overline n_{\rm in}=\langle0|b^\dagger b|0\rangle =0 , 
\label{545}
\eeq
we end up with a final number of produced pairs which is nonzero, in
general, and is controlled by the Bogoliubov coefficient $c_-$ as 
\beq
\overline n_{\rm out}=\langle0|a^\dagger a|0\rangle =|c_-|^2 \not= 0 .
\label{546}
\eeq

In a second quantization approach, the amplification of perturbations
can thus be seen as a process of pair production from the vacuum (or from
any otherwise specified initial state), under the action of a
time-dependent external field (the gravi-dilaton background, in the 
string cosmology case). Equivalently, the process can be described as a
``squeezing" of the initial state \cite{54} (this description is useful to
evaluate the associated entropy production \cite{55}), or, in a
semiclassical language, as a ``parametric amplification" \cite{56} of the
wave function $\psi_k$, which is scattered by an effective potential
barrier through an ``anti-tunnelling" process \cite{57}. Quite
independently of the adopted language, the differential energy density of
the produced radiation, for each mode $k$, depends on the number of
produced pairs, and can be written as
\beq
d \r_k = 2 k\overline n_k {d^3 k \over (2\pi)^3}, ~~~~~~~~~~
\overline n_k = |c_-(k)|^2.
\label{547}
\eeq
The computation of the so-called energy spectrum, defined as the spectral
energy density per logarithmic interval of frequency, 
\beq
{d \r_k\over d \ln k}\equiv k {d \r_k\over d  k}=
{k^4\over \pi^2}|c_-(k)|^2, 
\label{547a}
\eeq
thus requires the computation of $c_-(k)$, and then the knowledge of the
asymptotic solution of the canonical pertubation equation at large
positive times. 

To give an explicit example we shall consider here a very simple
model consisting of two cosmological phases, an initial 
accelerated evolution up to the time $\eta_1$, and a
subsequent radiation-dominated evolution for $\eta >\eta_1$:
\bea
&&
a \sim (-\eta)^\a,  ~~~~~~~~~~~~~~~~\eta<\eta_1, \nonumber\\
&&
a \sim \eta,  ~~~~~~~~~~~~~~~~~~~~~~\eta>\eta_1.
\label{547b}
\eea
The effective potential for tensor perturbations in the E-frame, 
$|a''/a|$, starts from zero at $-\infty$, grows like $\eta^{-2}$, reaches a
maximum $\sim\eta_1^{-2}$, and vanishes in the radiation phase. We
must solve the canonical perturbation equation for $\eta <\eta_1$ and
$\eta >\eta_1$. In the first phase the equation 
reduces to a Bessel equation:
\beq
\psi_k^\se +\left[k^2-{\a(\a-1)\over \eta^2}\right]\psi_k=0,
\label{548}
\eeq
with general solution \cite{58}
\beq
\psi_k= |\eta|^{1/2}
\left[AH_\nu^{(2)}(|k\eta|)+BH_\nu^{(1)}(|k\eta|)\right], ~~~~~~
\nu=|\a-1/2|, 
\label{549}
\eeq
where $H_\nu^{(1,2)}$ are the first and second kind Hankel functions,
of argument  $k\eta$ and index $\nu=|\a-1/2|$ determined by the
kinematics of the background. By using the large argument limit 
of the Hankel functions for $\eta \ra -\infty$,
\beq
H_\nu^{(2)}(k\eta)\sim {1\over \sqrt{k\eta}}e^{-ik\eta}, ~~~~~~~~~~~~
H_\nu^{(1)}(k\eta)\sim {1\over \sqrt{k\eta}}e^{+ik\eta},
\label{550}
\eeq
we choose initially a positive frequency mode,  normalizing the solution to
a vacuum fluctuation spectrum, 
\beq
A=1/2, ~~~~~~~~~~~B=0.
\label{551}
\eeq
In the second phase $V=0$, and we have the  free oscillating
solution :
\beq
\psi_k= {1\over \sqrt{k}}\left(c_+ e^{-ik\eta}+c_- e^{+ik\eta}\right).
\label{552}
\eeq
The matching of $\psi$ and $\psi'$ at $\eta=\eta_1$ gives now the
coefficients $c_\pm$ (more precisely,
the matching would  require the continuity of the perturbed metric
projected on a spacelike hypersurface containing $\eta_1$, and the
continuity of the extrinsic curvature of that hypersurface \cite{59};  but
in many cases these conditions are equivalent to the continuity of the
canonical variable $\psi$, and of its first time derivative). 

For an approximate determination of the spectrum, which is often
sufficient for pratical purposes, it is convenient to distinguish two
regimes, in which the comoving  frequency $k$
is much higher or much lower than the frequency associated to the top of
the effective potential barrier, $|V(\eta_1)|^{1/2} \simeq \eta_1^{-1}$. 
In the first case, $k\gg {1/|\eta_1|}\equiv k_1$, we can approximate 
the Hankel functions with their large argument limit, and we find that 
there is no particle production,  
\beq
|c_+|\simeq 1, ~~~~~~~~~~
 |c_-|\simeq 0 .
\label{553}
\eeq
In practice, $c_-$ is not exactly zero, but is exponentially suppressed
as a function of the frequency, just like the quantum reflection probability
for a wave with a frequency well above the top of a potential step.  We
will neglect such an effect here, as we are mainly interested in a
qualitative estimate of the perturbation spectrum. 

In the second case,  $k\ll {1/|\eta_1|}\equiv k_1$, we can use the small
argument limit of the Hankel functions,
\beq
H_\nu^{(2)} \sim a (k\eta_1)^\nu -i b  (k\eta_1)^{-\nu}, ~~~~~~~~~~
H_\nu^{(1)} \sim a (k\eta_1)^\nu +i b  (k\eta_1)^{-\nu}, 
\label{554}
\eeq
and we find
\beq
|c_+|\simeq 
 |c_-|\simeq |k\eta_1|^{-\nu-1/2},
\label{555}
\eeq
corresponding to a power-law spectrum:
\beq
{d \r_k\over d \ln k}=
{k^4\over \pi^2}|c_-(k)|^2\simeq {k_1^4\over \pi^2}
\left(k\over k_1\right)^{3-2\nu}, ~~~~~~~~~~~~k<k_1,
\label{556}
\eeq
with a cut-off frequency $k_1=\eta_1^{-1}$ controlled by the height of the
effective potential. 

For a comparison with present observations, it is finally convenient to
express the spectrum in terms of the proper frequency, $\om(t)=k/a(t)$,
and in units of critical energy density, $\r_c(t)=3M_p^2H^2(t)/8\pi$. We
then obtain the dimensionless spectral distribution, 
\bea
\Om(\om, t)={\om \over \r_c(t)}{d\r (\om)\over d\om} &\simeq &{8\over
3\pi}{\om_1^4\over M_p^2H^2}
\left(\om\over \om_1\right)^{3-2\nu} \nonumber\\
&\simeq&
g_1^2 \Om_\ga(t)
\left(\om\over \om_1\right)^{3-2\nu}, 
~~~~~~~~~\om<\om_1, 
\label{557}
\eea
where
\beq
\om_1={k_1\over a} ={1\over a\eta_1}\simeq {H_1a_1\over a}
\label{558}
\eeq
is the maximal amplified proper frequency, $g_1= H_1/M_p$, and 
\beq
\Om_\ga(t)={\r_\ga\over \r_c}=
\left(H_1\over H\right)^2\left(a_1\over a\right)^4
\label{559}
\eeq
is the energy density (in critical units) of the radiation that becomes
dominant at $t=t_1$, rescaled down at a generic time $t$ (today, $\Om_\ga
(t_0) \sim 10^{-4}$). 

It is important to stress that the amplitude of the spectrum is controlled
by $g_1= H_1/M_p$, i.e. by the curvature scale in Planck units at the time
of the transition $t_1$ (a fundamental parameter of the given inflationary
model). The slope of the spectrum, $3-2\nu$, is instead controlled by the
kinematics  of the background. In fact, it depends on the Bessel index
$\nu$ which, in its turn, depends on $\a$, the power of the scale factor
(eq. (\ref{549}). The behaviour in frequency of the graviton spectrum, in
particular, tends to follow the behaviour of the curvature scale during the
epoch of accelerated evolution, see Table III.

\begin{table}
\tabcolsep .1cm
\renewcommand{\arraystretch}{2.0}
\begin{center}
\begin{tabular}{|c||c||c||c|}
\hline
   & scale factor & Bessel index &  spectrum  \\ 
\hline
de Sitter, constant curvature &$\a=-1 $   &   $3-2\nu =0$   & flat \\ 
\hline
power-inflation, decreasing curvature 
& $\a<-1  $ &  $3-2\nu<0$  & decreasing\\ 
\hline
pre-big bang inflation, growing curvature 
& $\a>-1 $    &  $3-2\nu >0$  & increasing\\ 
\hline 
\end{tabular}
\bigskip
\caption{Slope of the graviton spectrum.}
\end{center}
\end{table}

The standard inflationary scenario is thus characterized by a flat or
decreasing graviton spectrum; in  string cosmology,  instead, we must
expect  a growing spectrum. This has important phenomenological
implications, that will be discussed in the following section. 

\section{The relic graviton background}
\label{VI}

As discussed in the previous Section, one of the most firm predictions of
all inflationary models is the amplification of the traceless, transverse
part of the quantum fluctuations of the metric tensor, and the formation
of a primordial, stochastic background of relic gravitational waves,
distributed over a quite large range of frequencies (see \cite{60} for a
discussion of the stochastic properties of such a background, and \cite{61}
for a possible detection of the associated ``squeezing" \cite{62}). 

In a string cosmology context, the expected graviton background has been
already discussed  in a number of detailed review papers \cite{57,63,64}.
Here I will summarize the main properties of the background predicted in
the context of the pre-big bang scenario. 

For a phenomelogical discussion of the spetrum, it is convenient to
consider the plane $\{ \Om_G, \om\}$. In this plane there are three main
phenomenological constraints:

\begin{itemize}

\item{}A first constraint comes from the large scale isotropy of the CMB
radiation. The degree of anisotropy measured by COBE imposes a bound on
the energy density of the graviton background at the scale of the present
Hubble radius \cite{65}, 
\beq
\Om_G(\om_0)~ \laq~ 10^{-10}, ~~~~~~~~~~~~~~~~~
\om_0 \sim 10^{-18}~ {\rm Hz}.
\label{61}
\eeq

\item{}A second constraint comes from the absence of distortion of the
pulsar timing-data \cite{66}, and gives the bound
\beq
\Om_G(\om_p)~ \laq~ 10^{-8}, ~~~~~~~~~~~~~~~~~
\om_p \sim 10^{-8}~ {\rm Hz}.
\label{62}
\eeq

\item{}A third constraint comes from nucleosynthesys \cite{67}, which
implies that the total graviton energy density, integrated over all modes,
cannot exceed the energy density of one massless degree of freedom in
thermal equilibrium, evaluated at the nucleosynthesis epoch. This gives a
bound for the peak value of the spectrum \cite{33}, 
\bea
&&
h_{100}\int d \ln \om \Om_G (\om,t_0) ~\laq~ 0.5 \times 10^{-5}, 
\nonumber \\
&&
h_{100}= H_0/(100 ~{\rm km~sec}^{-1}{\rm Mpc}^{-1}), 
\label{63}
\eea
which applies to all scales.

\end{itemize}

A further bound can be obtained  by
considering the production of primordial black holes \cite{68}. The
production of gravitons, in fact, could be associated to the formation
of black holes, whose possible evaporation, at the present epoch, is
constrained by a number of astrophysical observations. The absence of
evaporation imposes an indirect upper limit on the graviton background.  
In a string cosmology context, however, and in the frequency range
of interest for observations,  this upper limit is roughly of the same order
as the nucleosynthesis bound \cite{68}. 

For flat or decreasing spectra it is now evident that the more constraining
bound is the low-frequency one, obtained from the COBE data. In the
standard inflationary scenario,  characterized by flat or decreasing
spectra (see Table III), the maximal allowed graviton background can thus
be plotted as in Fig. 8. The flat spectrum corresponds to de Sitter inflation,
the decreasing spectra to power inflation. The breakdown in the spectrum,
around $\om_eq \sim 10^{-16}$ Hz, is due to the transition from the
radiation-dominated to the matter-dominated phase, which only affects
the low-frequency part of the spectrum, namely those modes re-entering
the horizon in the matter-dominated era. For such modes there is an
additional potential barrier in the canonical perturbation equation, which
induces an additional amplification $\sim (\om_eq/\om)^2 >1$, with
respect to the flat de Sitter spectrum. 

\begin{figure}
\centerline{\epsfxsize=8.0cm
\epsffile{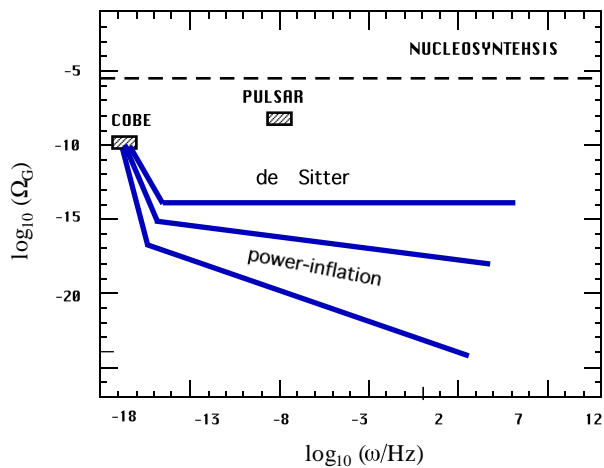}}
\caption{Graviton spectra in the standard inflationary scenario.} 
\label{fig8}
\end{figure}

However, this break of the spectrum is not important for our purposes.
What is important is the fact that the observed anisotropy constrains the
maximal amplitude of the spectrum. But the amplitude depends on the
inflation scale, as stressed in the previous Section. From the COBE bound
(\ref{61}), imposed  on the modified de Sitter spectrum at the Hubble
scale, 
\beq
\Om_G (\om_0,t_0)= g_1^2 \Om_\ga (t) (\om_eq/\om)^2 ~\laq~ 10^{-10}, 
\label{64}
\eeq
we thus obtain a direct constraint on the inflation scale:
\beq
H_1/M_p ~ \laq ~ 10^{-5}
\label{65}
\eeq
(for power-inflation the bound is even stronger \cite{31}). 

This bound applies to all models characterized by a flat or decreasing
spectrum. The bound can be evaded, however, if the spectrum is growing,
like in the string cosmology context. To illustrate this point, let me consider
the simplest class of the so-called ``minimal" pre-big bang models,
characterized by three main kinematic phases \cite{32,44}: an
initial low-energy, dilaton-driven phase,  an intermediate 
high-energy ``string" phase, in which $\ap$ and loop corrections become
important, and a final standard, radiation-dominated phase (see
Fig. 9). The time scale $\eta_s$ marks the transition to the high curvature
phase, and the time scale $\eta_1$, characterized by a final curvature of
order one in string units, marks the transition to the radiation-dominated
cosmology. 

\begin{figure}
\centerline{\epsfxsize=8.0cm
\epsffile{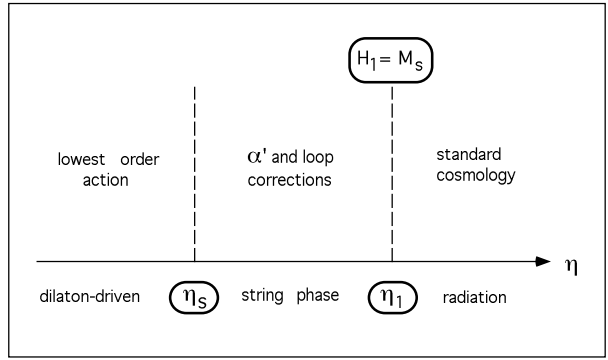}}
\caption{A minimal model of pre-big bang background.} 
\label{fig9}
\end{figure}

By computing graviton production in this background \cite{32} we find
that the spectrum is characterized by two branches: a high frequency
branch, for modes crossing the horizon (or ``hitting" the barrier) in the
string phase, $\om >\om_s=(a\eta_s)^{-1}$; and a low frequency
branch, for modes crossing the horizon in the
initial dilaton phase, $\om <\om_s$. The slope is cubic at low frequency,
and flatter at high frequency, and the spectrum can be parametrized as
follows:
\bea
\Om_G(\om,t_o) &\simeq & g_1^2
\Om_\ga (t_0) \left(\om\over \om_{1}\right)^{3-2\nu},
~~~~~~~~~~~~~~~~~~~~~~~~~\om_{s}<\om<\om_1,
\nonumber\\ 
& \simeq & g_1^2
\Om_\ga(t_0) \left(\om_1\over \om_{s}\right)^{2\nu}
\left(\om\over \om_{1}\right)^{3}, 
~~~~~~~~~~~~~~~~~~~\om<\om_{s} 
\label{66}
\eea
(modulo logarithmic corrections).  There are two main parameters: the
transition frequency $\om_s$, and the Bessel index $\nu$, for the high
frequency part of the spectrum. These parameters represent our present
ignorance about the duration and the kinematic details of the high
curvature phase. 

In spite of this uncertainty, however, there is a rather precise prediction
for the height and the position of the peak of the spectrum, which turns
out to be fixed in terms of the fundamental ratio $M_s/M_p$ as:
\beq
\Om_G(\om_1) \sim 10^{-4}(M_s/M_p)^2, ~~~~~~~~~~~~~~~
\om_1 \sim 10^{11}(M_s/M_p)^{1/2} ~{\rm Hz} .
\label{67}
\eeq
The behaviour of the spectrum, in this class of models, is illustrated in Fig.
10. A precise computation \cite{33} shows that, given the maximal
expected value of the string scale \cite{13} ($M_s/M_p\simeq 0.1$), the
peak value is automatically compatible with the nucleosynthesis bound, as
well as with bounds from the production of primordial black holes.  

In the minimal models the position of the peak is fixed. Actually, what is
really fixed, in string cosmology, is the maximal height of the peak, but
not necessarily the position in frequency, and it is not impossible, in more
complicated non-minimal models, to shift the peak at lower frequencies. 

\begin{figure}
\centerline{\epsfxsize=8.0cm
\epsffile{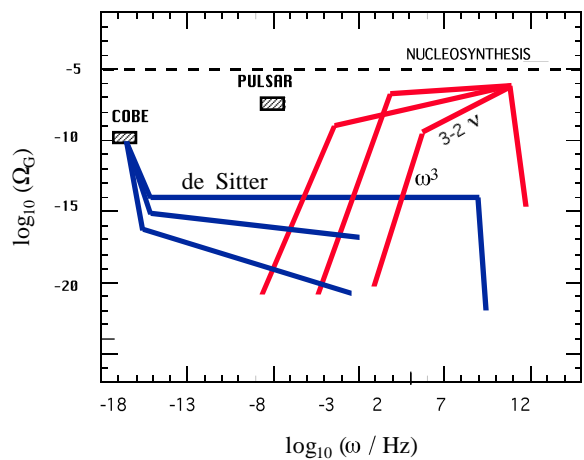}}
\caption{Graviton spectra in minimal  pre-big bang models.} 
\label{fig10}
\end{figure}

In minimal models, in fact, the  beginning of the radiation phase coincides
with the end of the string phase. We may also consider models, however, 
in which the dilaton coupling $g_s^2= \exp (\phi)$ is still small at the
end of the high curvature  phase, and the radiation era begins much later,
when $g_s^2 \sim  1$, after an intermediate dilaton-dominated regime.
The main difference between the two cases is that in the second case the
effective potential which amplifies tensor perturbations is 
non-monotonic \cite{57}, so that there are high frequency modes
re-entering the horizon before the radiation era. As a consequence, the
perturbation spectrum is also non-monotonic, and the peak does not
coincide any longer, in general,  with the end point of the spectrum (see
also \cite{69}), as illustrated in Fig. 11.

For minimal models the peak is around $100$ GHz, for non minimal models
it could be at lower frequencies. These are good news from an
experimental point of view, of course, but non-minimal models seem to be
less natural, at least from a theoretical point of view. The boxes around
the peak, appearing in Fig. 11, represent the uncertainty in the position of
the peak due to our present ignorance about the precise value of the ratio
$M_s/M_p$ (for the illustrative purpose of Fig. 11, the ratio is assumed to
vary in the range $0.1- 0.01$). The wavy line, in the high frequency branch
of the spectrum, represents the fact that the spectrum associated to the
string phase could be monotonic on the average, but locally oscillating
\cite{70}. Finally, the lower strip, labelled by $\da  s= 99 \%$, represents
the fact that even the height of the peak could be lower than expected, if
the produced gravitons have been diluted by some additional reheating
phase, occurring well below the string scale, during the standard
evolution. 

This last effect can be parameterized in terms of $\da s$, which is the
fraction of present entropy density in radiation, due to such additional,
low-scale reheating. The position of the peak then depends on $\da
s$ as \cite{33}: 
\bea
&&
\om_1(t_0) \simeq  T_0\left(M_s\over
M_P\right)^{1/2}\left(1-\da s\right)^{1/3} , \nonumber \\
&&
\Om_G(\om_1, t_0)\simeq 7\times 10^{-5}h_{100}^{-2}
\left(M_s\over M_P\right)^{2}
\left(1-\da s\right)^{4/3},
\label{68}
\eea
where $T_0 =2.7 {}^0K\simeq 3.6 \times 10^{11} {\rm Hz}$. Such a
dependence is not dramatic, however, because even for $\da  s= 99 \%$ the
peak keeps well above the standard inflationary prediction, represented
by the line labelled ``de Sitter" in Fig. 11. 

Given the various theoretical uncertainties, the best we can do, at
present, is to define the maximal allowed region for the expected
graviton background, i.e. the region spanned by the spectrum when all its
parameters are varied. Such a region is illustrated in Fig. 12, for the
phenomenologically interesting high frequency range. The figure
emphasizes the possible, large enhancement (of about eight orders of
magnitude) of the intensity of the  background in string cosmology, with
respect to the standard inflationary scenario. 

\begin{figure}
\centerline{\epsfxsize=8.0cm
\epsffile{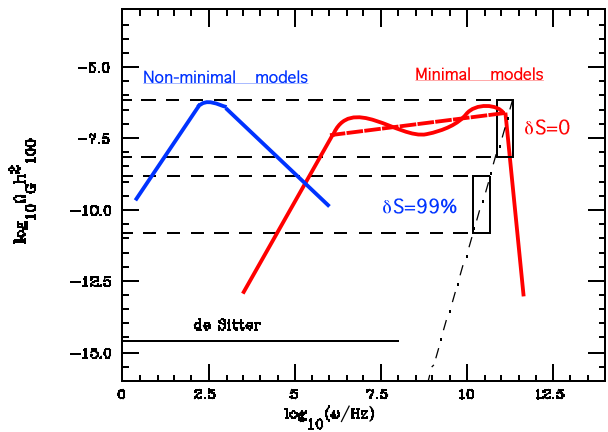}}
\caption{Peak and end point of the spectrum in minimal and non-minimal
models.}  
\label{fig11}
\end{figure}

It may be useful to stress again the reason of such enhancement. In the
standard inflationary scenario the graviton spectrum is decreasing, the
normalization is imposed at low frequency, and the peak value is
contolled by the anisotropy of the CMB radiation, $\da T/T ~\laq~10^{-5}$.
Thus, at low frequency,
\beq
\Om_G(t_0) ~\laq~ \Om_\ga (t_0)\left(\Da T\over T\right)^2_{COBE} \sim 
10^{-14}.
\label{69}
\eeq
In string cosmology the spectrum is growing, the normalization is imposed
at high frequency, and the peak value is controlled by the fundamental
ratio $M_s/M_p ~\laq~ 0.1$. Thus 
\beq
\Om_G(t_0) ~\laq~ \Om_\ga (t_0)\left(M_s\over M_p\right)^2 ~\laq  
~10^{-6}.
\label{610}
\eeq

The graviton background obtained from the amplification of the vacuum
fluctuations, in string cosmology and in standard inflation, is compared in
Fig. 12 with other, more unconventional graviton spectra. In particular: 
the graviton spectrum obtained 
from cosmic strings and topological defects \cite{71}, from bubble collision
at the end of a first order phase transition \cite{72}, and from a phase of
parametric resonance of the inflaton oscillations \cite{73}. Also shown in
Fig. 12 is the spectrum from models of quintessential inflation
\cite{74}, and a thermal black body spectrum for a temperature of about
one kelvin. All these cosmological backgrounds are higher than the
background expected from the  vacuum fluctuations in standard inflation,
but not in a string cosmology context.

It may be interesting, at this point, to recall the expected sensitivities of
the present, and near future, gravitational antennas, referred to the plots
of Fig. 12. 

At present, the best direct, experimental upper bound on the energy of a
stochastic graviton background comes from the cross-correlation of the
data of the two resonant bars NAUTILUS and EXPLORER \cite{75}: 
\beq
\Om_G h_{100} ~ \laq ~60, ~~~~~~~~~~~~~~~~~~~
\nu\simeq 907 ~{\rm Hz}
\label{611}
\eeq
(similar sensitivities are also reached by AURIGA \cite{76}).  
Unfortunately, the bound is too high to be significant for the plots of Fig.
12. However, a much better sensitivity, $\Om_G \sim
10^{-4}$ around  $\nu \sim 10^{3}$ Hz, is expected from the present
resonant bar detectors, if the integration time of the data is extended to
about one year. A similar, or slightly better sensitivity, 
$\Om_G \sim 10^{-5}$ around  $\nu \sim 10^{2}$ Hz, is expected from the
first operating version of the interferometric detectors, such as LIGO and
VIRGO. At high frequency,  from the kHz to the MHz range, a promising
possibility seems to be the use  of resonant electromagnetic cavities as
gravity wave detectors \cite{77}. Work is in progress \cite{78} to attempt
to improve their sensitivity. 

\begin{figure}
\centerline{\epsfxsize=8.0cm
\epsffile{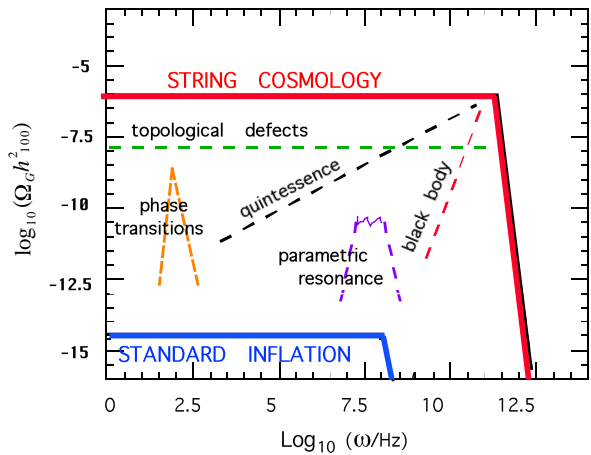}}
\caption{Allowed region for the spectrum of vacuum fluctuations in
string cosmology and in standard inflation, compared with other
relic spectra of primordial origin.}   
\label{fig12}
\end{figure}

The present, and near future, available sensitivities of resonant bars and
interferometers, therefore, are still outside the allowed region of Fig. 12,
determined by the border line
\beq
\Om_G h_{100} \simeq 10^{-6}.
\label{612}
\eeq
Such sensitivities are not so far from the border, after all, but to get
really inside we have to wait for the  cross-correlation of two spherical
resonant-mass detectors \cite{79},  expected to reach $\Om_G \sim
10^{-7}$ in the kHz range,   or for the advanced  interferometers,
expected to reach $\Om_G \sim 10^{-10}$ in the range of $10^{2}$ Hz. At
lower frequencies, around $10^{-2} - 10^{-3}$ Hz, the
space interferometer LISA \cite{36} seem to be able to reach very high
sensitivities,  up to $\Om_G \sim 10^{-11}$. Work is in progress, however,
for a more precise computation of their sensitivity to a cosmic stochastic 
background \cite{80}. 

Detectors able to reach, and to cross the limiting sensitivity (\ref{612}), 
could explore for the first time the parameter space of string cosmology
and of Planck scale physics. The detection of a signal from a pre-big bang
background, extrapolated to the GHz range, could give a first
experimental indication on the value of the fundamental ratio $M_s/M_p$.
Even the absence of a signal, inside the allowed region, would be
significant, as we could exclude some portion of parameter space of the
string cosmlogy models, obtaining in such a way  direct experimental
information about processes occuring at (or very near to) the string scale. 

\section{Conclusion}
\label{VII}

The conclusion of these lectures is very short  and simple. 

There is a rich structure of stochastic, gravitational wave backgrounds, of
cosmological origin, in the frequency range of present (or planned for the
future) gravity wave detectors. 

Among such backgrounds, the stronger one seems to be the background
possibly predicted in the context of the pre-big bang scenario, in a string
cosmology context, originating at (or very near to) the fundamental string
scale. Also, the maximal predicted intensity of the background seems to be
accessible to the sensitivity of the future advanced detectors. 

If this is the case, the future gravity wave detectors will be able to test
string theory models, or perhaps models referring to some more
fundamental unified theory, such as D-brane theory, M-theory, and so on.
In any case, such detectors will give direct experimental information on
Planck scale physics. 

\bigskip
\ack

It is a pleasure to thank the organizers of the {\sl Second SIGRAV School},
and all the staff of the Center  ``A. Volta", in Villa Olmo, for the pleasant
hospitality and the perfect organization of this interesting School. 

\bigskip

\appendix

\section{The string effective action}
\label{A}

The motion of a point particle in an external gravitational field,
$g_{\mu\nu}$,  is governed by the action
\beq
S= -{m\over 2} \int d\tau \dot x^\mu \dot x^\nu g_{\mu\nu}(x), 
\label {a1}
\eeq
where $x^\mu(\tau)$ are the spacetime coordinates of the particle, and
$\tau$ is an affine parameter along the particle world-line. 

The time evolution of a one-dimensional object like a string describes a
world-surface, or ``world-sheet", instead of a  world-line, and the action
governing its motion is given by the surface integral
\beq
S= -{M_s^2\over 2} \int d \tau d\sg \sqrt{-\ga}
\ga^{ij}\pa_ix^\mu\pa_jx^\nu g_{\mu\nu}(x),
\label{a2}
\eeq
where $\pa_i \equiv \pa/\pa \xi^i$, and  $\xi^i= (\tau,\sg)$  are
respectively the timelike and spacelike coordinate on the string
world-sheet ($i,j=1,2$). The coordinates $x^\mu(\tau,\sg)$ are the fields
governing the embedding of the string world-sheet in the external (also
called ``target") space. The parameter $M_s^2$ defines (in units
$h/2\pi=1=c$) the so-called string tension (the mass per unit length), and
its inverse defines the fundamental length scale of the theory (often
called, for hystorical reasons, the $\ap$ parameter):
\beq
M_s^2 \equiv {1\over \la_s^2} \equiv {1\over 2\pi \ap}. 
\label{a3}
\eeq
In a curved metric background $g_{\mu\nu}$ depends on $x^\mu$, and the
nonlinear action (\ref{a2}) represents what is called a ``sigma  model"
defined on the string world sheet. 

For the point particle action (\ref{a1}) the variation with respect to
$x^\mu$ leads to the well known geodesic equations of motion,
\beq
\ddot x^\mu +\Ga_{\a\b}\,^\mu \dot x^\a \dot x^\b =0.
\label{a4}
\eeq
The string equations of motion  are similarly obtained by varying with
respect to $x^\mu$ the action (\ref{a2}): we get then the Eulero-Lagrange
equations
\beq
\pa_i {\pa L\over \pa (\pa_ix^\mu)}= {\pa L\over \pa x^\mu}, ~~~~~~~~~~
L= \sqrt{-\ga}\ga^{ij}\pa_ix^\mu\pa_jx^\nu g_{\mu\nu}, 
\label{a5}
\eeq
which can be written explicitly as
\beq
\square x^\mu +\ga^{ij}\pa_ix^\a\pa_jx^\b \Ga_{\a\b}\,^\mu =0, 
~~~~~~~~~~~ \square  \equiv {1\over \sqrt{-\ga}} \pa_i 
\sqrt{-\ga}\ga^{ij}\pa_j ,
\label{a6}
\eeq
These equations describe the geodesic evolution of a test string in a
given external metric. The variation with respect to $\ga_{ik}$ imposes
the so-called ``costraints", i.e. the vanishing of the world sheet stress
tensor $T_{ik}$,
\beq
T_{ij}= {2\over \sqrt{-\ga}}{\da S \over \da \ga^{ij}} = 
\pa_ix^\mu\pa_jx^\nu g_{\mu\nu}- {1\over 2} \ga_{ij}
\pa_kx^\mu\pa^kx_\mu=0.
\label{a7}
\eeq

It is important to note, at this point, that for a classical string it is always
possible to impose the so-called ``conformal gauge" in which the world
sheet metric is flat, $\ga_{ij}=\eta_{ij}$. In fact, in an appropriate basis,
the two-dimensional metric tensor can   always be set in a diagonal form, 
$\ga_{ij}= {\rm diag} (a,b)$, and then, by using reparametrization
invariance on the world sheet, $b^2 d\sg^2= a^2 d\sg^{'2}$, 
the metric can be set in a
conformally flat form, $\ga_{ij}= a^2 \eta_{ij}$. Since the action (\ref{a2})
in invariant under the conformal (or Weyl) transformation $\ga_{ij} \ra
\Om^2 (\xi_k)\ga_{ij}$,
\beq
\sqrt{-\ga}\ga^{ij} \ra \sqrt{\Om^4}~ \Om^{-2} \sqrt{-\ga}\ga^{ij},
\label{a8}
\eeq
we can always eliminate the conformal factor $a^2$ in front of the
Minkowski metric, by choosing  $\Om= a^{-1}$. In the conformal gauge the
equations of motion (\ref{a6}) reduce to
\beq
\ddot x^\mu -x^{\prime\prime \mu} +\Ga_{\a\b}\,^\mu \left(\dot x^\a +
x^{\prime \a}\right) \left(\dot x^\b -x^{\prime \b}\right) =0,
\label{a9}
\eeq
where $\dot x= dx/d\tau$, $x' =dx /d\sg$, and the constraints (\ref{a7})
become 
\beq
g_{\mu\nu} \left(\dot x^\mu \dot x^\nu+x^{\prime \mu}
x^{\prime \nu}\right)=0, ~~~~~~~~~
g_{\mu\nu} \dot x^\mu x^{\prime \nu}=0.
\label{a10}
\eeq

We now come to the crucial observation which leads to the effective
action governing the motion of the background fields. The conformal
transformation (\ref{a8}) is an invariance of the classical theory. Let us
require that there are no ``anomalies", i.e. no quantum violations of this
classical symmetry. By imposing such a constraint, we will obtain a set of
differential equations to be satisfied by the background fields coupled to
the string. Thus, unlike a point particle which does not imposes any
constraint on the external geometry in which it is moving, the consistent
quantization of a string gives constraints for the external fields. The
background geometry cannot be chosen arbitrarily, but must satisfy the
set of equations(also called $\b$-function equations) which guarantee the
absence of conformal anomalies. The string effective action used in this
paper is the action which reproduces such a set of equations for the
background fields, and in particular for the metric. 

The derivation of the background equations of motion and of the effective
action, from the sigma-model action (\ref{a2}), can be performed order by
order by using a perturbative expansion in powers of $\ap$ (indeed, in the
limit $\ap \ra 0$ the action becomes very large in natural units, so that the
quantum corrections are expected to become smaller and smaller). Such a
procedure, however, is in general long and complicated, even to lowest
order, and a detailed derivation of the background equations is outside
the purpose of these lectures. Let me sketch here the procedure for the
simplest case in which the only external field coupled to the string is the
metric tensor $g_{\mu\nu}$. 

In the conformal gauge, the action (\ref{a2}) becomes:
\beq
S=-{1\over 4 \pi \ap} \int d^2 \xi~ \pa_i x^\mu\pa^i x^\nu g_{\mu\nu}. 
\label{a11}
\eeq
Let us formally assume a deformation of the number of world sheet
dimensions, from $2$ to $2+\ep$, and perform the conformal
transformation: $\eta_{ij} \ra \eta_{ij} \exp (\rho)$. Expanding we get,
for small $\ep$, 
\bea
&&
-{1\over 4 \pi \ap} \int d^{2+\ep}\xi ~ e^{\rho \ep /2}
\pa_i x^\mu\pa^i x^\nu g_{\mu\nu}= \nonumber \\
&&
-{1\over 4 \pi \ap} \int d^{2+\ep}\xi ~
\pa_i x^\mu\pa^i x^\nu g_{\mu\nu}\left(1+ {\ep \over 2} \rho+ ...\right) . 
\label{a12}
\eea

For $\ep \ra 0$ the $\r$ dependence disappears, and the classical 
action is conformally invariant. In order to preserve this invariance also
for the quantum theory, at the one loop level, let us treat the sigma model 
as a quantum field theory for $x^\mu(\sg, \tau)$, and let us
consider the quantum fluctuations $\hat x^\mu$ around a given
expectation value $x_0^\mu$.  For the general reparametrization
invariance of the theory we can always choose for $x_0$ a locally inertial
frame, such that $g_{\mu\nu}(x_0)=\eta_{\mu\nu}$. By expanding the
metric around $x_0$,  the leading corrections are of second order in the
fluctuations, because in a locally inertial frame the first derivatives of the
metric (and then the Cristoffel connection) can always be set to zero (but
not the curvature). With an appropriate choice of coordinates, called
Riemann normal coordinates, the metric can thus be expanded as:
\beq
g_{\mu\nu}(x)= \eta_{\mu\nu}- {1\over 3} R_{\mu\nu\a\b}(x_0)
\hat x^\a \hat x^\b + ...
\label{a13}
\eeq
and the action for the quantum fluctuations becomes, to lowest order in
the curvature,
\bea
&&\qquad
S=-{1\over 4 \pi \ap} \int d^{2+\ep}\xi \Bigg[
\pa_i \hat x^\mu\pa^i \hat x_\mu\left(1+ {\ep \over
2}\rho\right)\nonumber \\
 &&\qquad
 -{1\over 3} \pa_i \hat x^\mu\pa^i \hat x^\nu
R_{\mu\nu\a\b}(x_0) \hat x^\a \hat x^\b  \left(1+ {\ep \over 2}
\rho\right)+ ... \Bigg].  \label{a14}
\eea

It must be noted  that, at the quantum level, 
the dependence of $\r$ does not disappear in general from the action
 in the limit $\ep \ra 0$, since 
there are one-loop terms that diverge like $\ep^{-1}$, just to cancel the
$\ep$ dependence and to give a contribution proportional to $\r$ to the
effective action. By evaluating, for instance, the two point function for
the quantum operator $\hat x^\a \hat x^\b $, in the coincidence limit $\sg
\ra \sg'$ (the tadpole graph), one obtains \cite{81}
\beq
\langle \hat x^\a (\sg)\hat x^\b (\sg') \rangle_{\sg \ra \sg'} 
\sim \eta^{\a\b} \lim_{\sg \ra \sg'}\int d^{2+\ep} k {e^{ik \cdot (\sg
-\sg')}\over k^2}
 \sim \eta^{\a\b}  \ep^{-1},
\label{a15}
\eeq
which gives the one-loop contribution to the action
\beq
\Da S\sim \int d^{2+\ep}\xi~ \pa_i \hat x^\mu\pa^i \hat x^\nu
R_{\mu\nu} ~\r.
\label{a16}
\eeq
This term violates, at one-loop, the conformal invariance, unless we
restrict to a background geometry satisfying the condition
\beq
R_{\mu\nu}=0,
\label{a17}
\eeq
which are just the usual Einstein equations in vacuum. 

A similar procedure can be applied if the string moves in a richer external
background (not only pure gravity). Indeed, pure gravity is not enough,  as
a consistent quantum theory for closed bosonic strings, for instance, must
contain at least three massless state (beside the unphysical tachyon,
removed by supersymmetry) in the lowest energy level: the graviton, the
scalar dilaton and the pseudo-scalar Kalb-Ramond axion. The sigma model
describing the propagation of a string in such a background must thus
contain the coupling to the metric, to the dilaton $\phi$, and to the
two-form $B_{\mu\nu}=-B_{\nu\mu}$: 
\bea
&&
S= -{1\over 4 \pi \ap} \int d^{2}\xi~ 
\pa_i x^\mu\pa_j x^\nu \left(\sqrt{-\ga}\ga^{ij} g_{\mu\nu}+
\ep ^{ij} B_{\mu\nu}\right)\nonumber \\
&&
-{1\over 4 \pi} \int d^{2}\xi \sqrt{-\ga}{\phi\over 2}  R^{(2)}(\ga) , 
\label{a18}
\eea
where $\ep ^{ij}$ is the two-dimensional Levi-Civita densor density,
$\ep_{12}=-\ep_{21}=1$, and $R^{(2)}(\ga)$ is the two-dimensional scalar
curvature for the world sheet metric $\ga$. The condition of conformal
invariance, at the one-loop level, leads to the equations
\bea
&&
R_{\mu\nu} +\nabla_\mu \nabla_\nu-{1\over 4}H_{\mu\a\b}
H_\nu\,^{\a\b}=0, ~~~~
H_{\mu\nu\a}= \pa_\mu B_{\nu\a}+\pa_\nu B_{\a\mu}+\pa_\a B_{\mu\nu}, 
\nonumber \\
&&
R+ 2 \nabla^2 \phi -(\nabla\phi)^2 -{1\over 12} H_{\mu\nu\a}^2=0, 
\nonumber\\
&&
\nabla_\mu \left(e^{-\phi}H^{\mu\nu\a}\right)=0,
\label{a19}
\eea
which can be obtained by extremizing the effective action
\beq
 S= -{1\over 2\la_s^{d-1}}\int d^{d+1}x \sqrt{|g|}~ e^{-\phi}\left[R+
\left(\nabla\phi\right)^2- {1\over 12}H_{\mu\nu\a}^2
\right] .
\label{a20}
\eeq
(see Appendix C). 

It should be noted that the inclusion of the dilaton in the condition of
conformal invariance cannot be avoided, since the dilaton coupling in the
action (\ref{a18}) breaks conformal invariance already at the classical
level ($\sqrt{-\ga}  R^{(2)}$ is not invariant under a Weyl rescaling of
$\ga$). However, the dilaton term is of order $\ap$ with respect to the
other terms of the action (for dimensional reasons), so that it is correct to
sum up the classical dilaton contribution to the quantum, one-loop
effects, as they are all of the same order in $\ap$. Without the dilaton,
however, the world sheet curvature density $\sqrt{-\ga}  R^{(2)}$ does not
contribute to the string equations of motion, as it is a pure Eulero
two-form in two dimensions (just like the Gauss-Bonnet term in four
dimensions). 

Let me note, finally, that the expansion around $x_0$ can be continued to
higher orders, 
\beq
g(x)= \eta +R \hat x \hat x + \pa R \hat x\hat x\hat x+ 
R^2 \hat x\hat x\hat x\hat x + ... ,
\label{a21}
\eeq
thus introducing higher curvature terms, and higher powers of $\ap$, in
the effective action: 
\beq
S= -{1\over 2\la_s^{d-1}}\int d^{d+1}x \sqrt{|g|}~ e^{-\phi}\left[R+
\left(\nabla\phi\right)^2-{\ap\over 4} R_{\mu\nu\a\b}^2 + ...\right].
\label{a22}
\eeq
At any given order, unfortunately, there is an intrinsic ambiguity in the
action due to the fact that, with an appropriate field redefinition of order
$\ap$, 
\bea
&&
g_{\mu\nu} \ra g_{\mu\nu} +\ap\left(R_{\mu\nu}+ \pa_\mu\phi
\pa_\nu\phi+ ...\right) \nonumber \\
&&
\phi \ra \phi + \ap\left(R+ \nabla^2 \phi + ...\right),
\label{a23}
\eea
we obtain a number of different actions, again of the same order in $\ap$ 
(see for instance \cite{82}). This ambiguity cannot be eliminated until we
limit to an effective action truncated to a given finite order. 

The higher curvature (or higher derivative) expansion of the effective
action is typical of string theory: it is controlled by the fundamental,
minimal length parameter $\la_s = (2 \pi \ap)^{1/2}$, in such a way that
the higher order corrections disappear in the point-particle limit $\la_s
\ra 0$. At any given order in $\ap$, however, there is also the more
conventional expansion in power of the coupling constant $g_s$ (i.e. the
loop expansion of quantum field theory: tree-level $\sim g^{-2}$, one-loop 
 $\sim g^{0}$, two-loop  $\sim g^{2}$, ...). The important observation is
that, in a string theory context, the effective couplig constant is controlled
by the dilaton. Consider, for instance, a process of graviton scattering, in
four dimensions. Comparing the action (\ref{24}) with the standard,
gravitational Einstein action (\ref{21}), it follows that the effective
coupling constant, to lowest order, is
\beq
\sqrt{8\pi G}=\la_p= \la_s e^{\phi/2}
\label{a25}
\eeq
($G$ is the usual Newton constant). Each loop adds an integer power of the
square of the dimensionless coupling constant, which is controlled by
the dilaton as  \beq
g_s^2 =(\la_p/\la_s)^2 = (M_s/M_p)^2 = e^\phi.
\label{a26}
\eeq
We may thus expect, for the loop expansion of the action, the following
general scheme:
\bea
S=&-&\int e^{-\phi}\sqrt{-g}\left( R + \nabla \phi ^2 +\ap R^2+ ...\right)
~~~~~~~~~~~~~ {\rm tree~level}\nonumber \\
&-&\int \sqrt{-g}\left( R + \nabla \phi ^2 +\ap R^2+ ...\right)
~~~~~~~~~~~~~~~~~~ {\rm one-loop}\nonumber \\
&-&\int e^{+\phi}\sqrt{-g}\left( R + \nabla \phi ^2 +\ap R^2+ ...\right)
~~~~~~~~~~~~~ {\rm two-loop}\nonumber \\
&&
............................
\label{a27}
\eea
Unfortunately, each term in the action, at each loop order, is multiplied 
by a dilaton ``form factor" which is different in general for 
{\em different fields} and for {\em different orders}. This difference can
lead to an effective violation of the universality of the
gravitational interactions \cite{83} in the low-energy, macroscopic
regime,  and this violation can be reconciled with the present tests of the
equivalence principle only if the dilaton is massive enough, to make short
enough the range of the non-universal dilatonic interactions. 

The tree-level relation (\ref{a26}) is valid also for a higher-dimensional
effective action, provided $e^{\phi}$ represents the shifted
four-dimensional dilaton which includes the volume of the
extra-dimensional, compact internal space, and which controls the
grand-unification gauge coupling, $\a_{GUT}$, as \cite{13}
\beq
\a_{GUT}= \exp \langle \phi \rangle = (M_s/M_p)^2 \sim 0.1 - 0.001.
\label{a28}
\eeq
However, the relation  (\ref{a26}) is no longer valid, in general, 
if the gauge interactions are confined in four-dimensions and only gravity
propagates in the extra dimensions. In that case the relation depends on
the volume of the extra dimensions, whose size may be allowed to be large
in Planck units \cite{84}. For  internal dimensions of volume $V_n$
the relation becomes, in particular,
\beq
M_p^2= M_s^{2+n} V_n e^{-\Phi},
\label{a29}
\eeq
where $\Phi$ is the dilaton in $d=3+n$ dimensions. In this case, the
string mass parameter could  be much smaller than the value expected
from eq. (\ref{a28}), provided the internal volume is correspondingly
larger. 

\section{Duality symmetry}
\label{B}

{\underline {\bf Notations and conventions}}. In this paper we use the
metric signature $(+---)$, and we define the Riemann and Ricci tensor as
follows:
\bea
&&
R_{\mu\nu\a}\,^\b= \pa_\mu\Ga_{\nu\a}\,^\b+ \Ga_{\mu\r}\,^\b
\Ga_{\nu\a}\,^\r - (\mu \leftrightarrow \nu),
\nonumber\\
&&
R_{\nu\a}= R_{\mu\nu\a}\,^\mu.
\nonumber
\eea

Consider the gravi-dilaton effective action, in the S-frame, to lowest
order in  $\ap$ and in the quantum loop expansion: 
\beq 
S= -{1\over 2\la_s^{d-1}} \int d^{d+1}x \sqrt{|g|}~
e^{-\phi}\left[R+ \left(\nabla \phi\right)^2\right]. 
\label{b1}
\eeq
For a homogeneous, but anisotropic, Bianchi I type metric background:
\beq
\phi=\phi(t), ~~~~~~~ g_{00}= N^2(t), ~~~~~~~
g_{ij}=- a_i^2(t) \da_{ij},
\label{b2}
\eeq
we have ($H_i =\dot a_i/a_i$):
\bea
&&
\left(\nabla \phi \right)^2 = {\dot \phi^2\over N^2}, ~~~~~~~~~~ 
\sqrt{-g}= N \prod_{i=1}^d a_i,  \nonumber \\
&&
\Ga_{00}\,^0={\dot N\over
N}\equiv F, ~~~~~~~~~~ \Ga_{0i}\,^j= H_i \da_i^j, ~~~~~~~~~~
\Ga_{ij}\,^0= {a_i\dot a_i\over N^2}\da_{ij}, \nonumber \\
&&
R= {1\over N^2} \left[ 2 F \sum_i H_i -2 \sum_i \dot H_i -\sum_i H_i^2-
\left(\sum_i H_i\right)^2\right].
\label{b3}
\eea
By noting that 
\bea
&&
{d\over dt} \left[2{e^{-\phi}\over N}\left(\prod_{k=1}^d a_k \right)
\left(\sum_i H_i \right) \right]\nonumber \\
&&
= {e^{-\phi}\over N}\left(\prod_{k=1}^d a_k \right)
\left[2 \sum_i \dot H_i-2 
 F \sum_i H_i -2 \dot \phi \sum_i H_i +2\left(\sum_i H_i\right)^2 \right],
\label{b4}
\eea
the action (\ref{b1}), modulo a total derivative, can be rewritten as:
\beq
\fl 
S= -{1\over 2\la_s^{d-1}} \int d^{d}x dt \prod_{i=1}^d a_i
{e^{-\phi}\over N}\left[\dot \phi^2 - \sum_i H_i^2 +  \left(\sum_i
H_i\right)^2 -2 \dot \phi \sum_i H_i \right]. 
\label{b5}
\eeq

We now introduce the so-called shifted dilaton $\fb$, defined by
\beq
e^{-\fb}= \int{ d^{d}x \over \la_s^d}\prod_{i=1}^d a_i e^{-\phi},
\label{b6}
\eeq
from which
\beq
\phi= \fb + \sum_i \ln a_i, ~~~~~~~~~~~~
\dot \phi = \fbp + \sum_i H_i
\label{b7}
\eeq
(by assuming spatial sections of finite volume, $(\int
d^dx\sqrt{|g|})_{t={\rm const}}<\infty$,   we have absorbed
into $\phi$ the constant shift $-\ln(\la_s^{-d}\int  d^dx)$, required to
secure the scalar behaviour of $\fb$ under coordinate 
reparametrizations preserving the comoving gauge). The action becomes:
\beq
S= - {\la_s\over 2} \int dt {e^{-\fb}\over N} \left (\fbp^2 - \sum_i H_i^2
\right).
\label{b8}
\eeq
By inverting one of the $d$ scale factors the corresponding Hubble
parameter changes sign, 
\beq
a_i \ra \ti a_i = a_i^{-1}, ~~~~~~~~~
H_i \ra \ti H_i = {\dot {\ti a_i}\over  \ti a_i} =a_i {d a_i^{-1}\over dt} = -H_i,
\label{b9}
\eeq
so that the quadratic action (\ref{b8}) is clearly invariant under the
inversion of any scale factor preserving the shifted dilaton,
\beq
a_i \ra \ti a_i = a_i^{-1}, ~~~~~~~~~~~~~~
\fb \ra \fb 
\label{b10}
\eeq
(``scale factor duality", see \cite{9} and the first paper of Ref. \cite{8}). 

In order to derive the field equations, it is convenient to use the variables
$\b_i =\ln a_i$, so that $H_i= \dot \b_i$, $\dot H_i= \ddot \b_i$, and the
action (\ref{b8}) is cyclic in $\b_i$. By varying with respect to $N, \b_i$
and $\fb$, and subsequently fixing the cosmic time gauge $N=1$, we
obtain, respectively,
\bea
&&
\fbp^2 - \sum_i H_i^2=0,
\label{b11}\\
&&
\dot H_i -H_i \fbp=0,
\label{b12}\\
&&
2 \ddot {\fb} -\fbp^2 -\sum_i H_i^2=0.
\label{b13}
\eea
This is a system of $(d+2)$ equations for the $(d+1)$ variables $\{a_i,
\phi\}$. However, only $(d+1)$ equations are independent (see for instance
\cite{21}: eq. (\ref{b11}) represents a constraint on the set of initial data). 

The above equations are invariant under a time reversal transformation
\beq
t \ra -t, ~~~~~~~~~~ H \ra -H,  ~~~~~~~~~~\fbp \ra -\fbp, 
\label{b14}
\eeq
and also under the duality transformation (\ref{b10}). If we invert $k \leq
d$ scale factors, 
$\ti a_1 = a_1^{-1}$, ... $\ti a_k = a_k^{-1}$, the shifted dilaton is
preserved, $\fb = \ti {\fb}$, provided
\beq
\fb = \phi - \sum_{i=1}^d \ln a_i = \ti \phi -\sum_{i=1}^k \ln \ti a_i-
\sum_{i=k+1}^d \ln a_i, 
\label{b15}
\eeq
from which:
\beq
\ti \phi = \phi -2 \sum_{i=1}^k \ln  a_i.
\label{b16}
\eeq
Given an exact solution, represented by the set of variables
\beq
\{a_1, ...  a_d, ~\phi\},
\label{b17}
\eeq
the inversion of $k \leq d$ scale factors defines then a new exact
solution, represented by the set of variables
\beq
\{a_1^{-1}, ...  a_k^{-1}, a_{k+1}, ... a_d, ~\phi- 2 \ln a_1 
... -2 \ln a_k \}. 
\label{b18}
\eeq
By inverting all the scale factors we obtain the transformation 
\beq
\{a_i, \phi\} \ra \{a_i^{-1}, \phi- 2\sum _{i=1}^d \ln a_i \} 
\label{b19}
\eeq
which, in the isotropic case, corresponds in particular to the duality
transformation (\ref{25}). 

As a simple example, we consider here the particular isotropic solution
\beq
a= t^{1/\sqrt d}, ~~~~~~~~~~~~~~~
\fb = -\ln t,
\label{b20}
\eeq
which satisfies identically the set of equations (\ref{b11}) - (\ref{b13}). By
applying a duality {\em and} a time-reversal transformation we obtain the
four different exact solutions
\bea
&&
\{ a_{\pm} (t) = t^{\pm 1/\sqrt d},~~ \fb (t)= -\ln t \}, \nonumber \\
&&
\{ a_{\pm} (-t) = (-t)^{\pm 1/\sqrt d}, ~~\fb (-t)= -\ln (-t) \},
\label{b21}
\eea
corresponding to the four branches illustrated in Fig. 2, and describing
decelerated expansion, $a_+(t)$, 
decelerated contraction, $a_-(t)$, accelerated contraction, $a_+(-t)$, 
accelerated expansion, $a_-(-t)$. The solution describes expansion or
contraction if the sign of $\dot a$ is positive or negative, repectively, and
the solution is accelerated or decelerated if $\dot a$ and $\ddot a$ have
the same or the opposite sign, respectively. 

It is important to consider also the dilaton behaviour. According to eq.
(\ref{b7}): 
\beq
\phi_\pm (\pm t) =\fb (\pm t) + d \ln a_\pm (\pm t) =
\left( \pm \sqrt d -1\right) \ln (\pm t). 
\label{b22}
\eeq
It follows that, in a phase of growing curvature ($t<0, t \ra 0_-$), the
dilaton is growing only for an expanding metric, $a_-(-t)$. This means
that, in the isotropic case, there are only expanding pre-big bang
solution, i.e. solutions evolving from the string perturbative vacuum ($ H
\ra 0, \phi \ra -\infty$), and then characterized by a growing string
coupling, $\dot g_s = (\exp \phi/2)\dot{}>0$. 

In the more general, anisotropic case, and in the presence of contracting
dimensions, a growing curvature solution is associated to a growing
dilaton only for a large enough number of contracting dimensions. To make
this point more precise, consider the particular, exact solution of eqs.  
(\ref{b11}) - (\ref{b13}) with $d$ expanding and $n$ contracting
dimensions, and scale factors $a(t)$ and $b(t)$, respectively:
\beq
a = (-t)^{-1/\sqrt{d+n}}, ~~~~~
b= (-t)^{1/\sqrt{d+n}}, ~~~~~\fb =-\ln (-t), ~~~~ t \ra 0_-. 
\label{b23}
\eeq
This gives, for the dilaton,
\beq
\phi= \fb + d \ln a +n\ln b ={n-d-\sqrt{d+n}\over \sqrt{d+n}} \ln (-t),
\label{b24}
\eeq
so that the dilaton is growing if 
\beq
d+\sqrt{d+n}>n.
\label{b25}
\eeq
For $n=6$, in particular, this condition requires $d>3$. This could represent
a potential difficulty for the pre-big bang scenario, which might be solved,
however, by quantum cosmology effects \cite{85}. 

The scale factor duality of the action (\ref{b8}) is in general broken by the
addition of a non-trivial dilaton potential (unless the potential depends on
the dilaton through $\fb$, of course). 
When the antisymmetric tensor $B_{\mu\nu}$ is included in the action,
however, the scale factor duality can be lifted to a larger group of global
symmetry transformations. To illustrate this important aspect of the
string cosmology equations, we will consider here a set of cosmological
baground fields $\{\phi, g_{\mu\nu}, B_{\mu\nu} \}$, for which a
syncronous frame exists where $g_{00}=1, g_{0i}=0, B_{0\mu}=0$, and all
the components $\phi,  g_{ij}, B_{ij}$ do not depend on the spatial
coordinates. 

Let us write the action
\beq 
S= -{1\over 2\la_s^{d-1}} \int d^{d+1}x \sqrt{|g|}~
e^{-\phi}\left[R+ \left(\nabla \phi\right)^2- {1\over 12}
H_{\mu\nu\a}^2
\right] 
\label{b26}
\eeq
directly in the syncronous gauge, as we are not interested in the field
equations, but only in the symmetries of the action. We set $g_{ij}=
-\ga_{ij}$ and we find, in this gauge,
\bea
&&
\Ga_{ij}\,^0= {1\over 2} \dot \ga_{ij}, ~~~~~~~~~~
\Ga_{0i}\,^j={1\over 2}g^{jk}\dot g_{ik}= {1\over 2} \left(g^{-1}\dot 
g\right)_i\,^j= \left(\ga^{-1}\dot \ga\right)_i\,^j
\nonumber \\
&&
R_0\,^0 = -{1\over 4} {\rm Tr }\left(\ga^{-1}\dot \ga\right)^2 
-{1\over 2} {\rm Tr }\left(\ga^{-1}\ddot \ga\right)
-{1\over 2} {\rm Tr } \left(\dot \ga^{-1}\dot \ga\right), \nonumber\\
&&
R_i\,^j =  -{1\over 2} \left(\ga^{-1}\ddot \ga\right)_i\,^j 
-{1\over 4} \left(\ga^{-1}\dot \ga\right)_i\,^j 
{\rm Tr } \left(\ga^{-1}\dot \ga\right) 
+{1\over 2} \left(\ga^{-1}\dot \ga \ga^{-1}\dot \ga\right)_i\,^j ,
\label{b27}
\eea
where
\beq
 {\rm Tr }\left(\ga^{-1}\dot \ga\right)= \left(\ga^{-1}\right)^{ij}
\dot \ga_{ji}= g^{ij} \dot g_{ji}, 
\label{b28}
\eeq
and so on [note also that    $\dot \ga^{-1}$  means   
$\left(\ga^{-1}\right)\dot{}$]. Similarly we find, for the antisymmetric
tensor, 
\bea
&&
H_{0ij}= \dot B_{ij}, ~~~~~~~~~~~~~
H^{0ij}= g^{ik}g^{jl} \dot B_{kl} =\left( \ga^{-1} \dot B \ga^{-1}\right)^{ij}, 
\nonumber\\
&&
H_{\mu\nu\a}H^{\mu\nu\a}= 3 H_{0ij}H^{0ij}=
- 3{\rm Tr }\left(\ga^{-1}\dot B\right)^2.
\label{b29}
\eea
Let us introduce the shifted dilaton, by absorbing into $\phi$ the spatial
volume, as before:
\beq
\sqrt{|\det g_{ij}|} e^{-\phi}= e^{-\fb},
\label{b30}
\eeq
from which
\beq
\fbp= \dot \phi -{1\over 2} {d\over dt} \ln \left( \det \ga\right) = 
\dot \phi -{1\over 2} {\rm Tr }\left(\ga^{-1}\dot \ga\right).
\label{b31}
\eeq
By collecting the various contributions from $\phi, R$ and $H^2$, the action
(\ref{b26}) can  be rewritten as:
\bea
&&\qquad
S=-{\la_s\over 2} \int dt e^{-\fb} \Bigg[ \fbp^2 + 
{1\over 4} {\rm Tr }\left(\ga^{-1}\dot \ga\right)^2
-{\rm Tr }\left(\ga^{-1}\ddot \ga\right)
\nonumber \\
&&\qquad
-{1\over 2} {\rm Tr } \left(\dot \ga^{-1}\dot \ga\right)
+\fbp {\rm Tr }\left(\ga^{-1}\dot \ga\right)
+{1\over 4} {\rm Tr }\left(\ga^{-1}\dot B\right)^2 \Bigg].
\label{b32}
\eea
We can now eliminate the second derivatives, and the mixed term 
($\sim \fbp \dot \ga$), by noting that 
\beq
{d \over dt} \left[ e^{-\fb} {\rm Tr }\left(\ga^{-1}\dot \ga\right)\right]=
e^{-\fb}\left[{\rm Tr }\left(\ga^{-1}\ddot \ga\right)
+ {\rm Tr } \left(\dot \ga^{-1}\dot \ga\right)
-\fbp {\rm Tr }\left(\ga^{-1}\dot \ga\right)\right].
\label{b33}
\eeq
Finally, by using the identity,
\beq
\left(\ga^{-1}\right)\dot{}= -\ga^{-1}\dot \ga \ga^{-1}
\label{b34}
\eeq
(following from $g^{-1} g=\ga^{-1} \ga=I$), we can rewrite the action in
quadratic form, modulo a total derivative, as
\beq
S=-{\la_s\over 2} \int dt e^{-\fb} \left[ \fbp^2 
-{1\over 4} {\rm Tr }\left(\ga^{-1}\dot \ga\right)^2
+{1\over 4} {\rm Tr }\left(\ga^{-1}\dot B\right)^2 \right].
\label{b35}
\eeq

This action can be set into a more compact form by using the $2d \times
2d$ matrix $M$, defined in terms of the spatial components of the metric
and of the antisymmetric tensor, 
\bea
&&
M=\pmatrix{G^{-1} & -G^{-1}B \cr
BG^{-1} & G-BG^{-1}B \cr}, \nonumber \\
&&
G= g_{ij}\equiv-\ga_{ij},  ~~~~~ G^{-1}\equiv g^{ij}, ~~~~~ B\equiv 
B_{ij}, 
\label{b36}
\eea
and using also the matrix $\eta$, representing the invariant metric of the
$O(d,d)$ group in the off-diagonal representation,
\beq
\eta=\pmatrix{0 & I \cr I & 0 \cr}
\label{b37}
\eeq
($I$ is the unit $d$-dimensional matrix). 
By computing $M\eta, \dot M \eta$ and $(\dot M \eta)^2$ we find, in fact,
\beq
\fl 
{\rm Tr } \left(\dot M \eta\right)^2= 
2  {\rm Tr }\left[\dot \ga^{-1}\dot\ga+  \left(\ga^{-1}\dot
B\right)^2\right]=
2  {\rm Tr }\left[-\left(\ga^{-1}\dot \ga\right)^2 +\left(\ga^{-1}\dot
B\right)^2\right], 
\label{b38}
\eeq
and the action becomes
\beq
S=-{\la_s\over 2} \int dt e^{-\fb} \left[ \fbp^2 
+{1\over 8} {\rm Tr }\left(\dot M\eta\right)^2\right].
\label{b39}
\eeq

We may note, at this point, that $M$ is  a symmetric matrix
of the pseudo-orthogonal  $O(d,d)$ group. In fact,
\beq
M^T \eta M =\eta, ~~~~~~~~~~ M=M^T
\label{b40}
\eeq
for any $B$ and $G$. Therefore:
\beq
M\eta=\eta M^{-1}, ~~~~~~~~\left(\dot M\eta\right)^2=
\eta \left( M^{-1}\right)\dot {} M \eta,
\label{b41}
\eeq
and the action can be finally rewritten as
\beq
S=-{\la_s\over 2} \int dt e^{-\fb} \left[ \fbp^2 
+{1\over 8} {\rm Tr } ~\dot M \left(M^{-1}\right)\dot{}\right].
\label{b42}
\eeq
This form is explicitly invariant under the global $O(d,d)$ transformations
(\ref{210}), preserving the shifted dilaton: 
\beq
\fb \ra \fb, ~~~~~~~~ M \ra \La^{T} M \La, ~~~~~~~~
\La^T\eta \La =\eta
\label{b43}
\eeq
In fact
\beq
{\rm Tr }~ \dot {\ti M }\left({\ti M}^{-1}\right)\dot{}=
{\rm Tr } \left [\La^T \dot M \La \La^{-1} \left(M^{-1}\right)\dot{}
\left(\La^{T}\right)^{-1} \right]=
 {\rm Tr } ~\dot M \left(M^{-1}\right)\dot{}.
\label{b44}
\eeq
In the absence of the antysimmetric tensor $M$ is diagonal, and the
special $O(d,d)$ transformation with $\La = \eta$ corresponds to an
inversion of the metric tensor: 
\bea
&&
M ={\rm diag} (G^{-1}, G), \nonumber \\
&&
\ti M = \La^{T} M \La= \eta M \eta = {\rm diag} (G, G^{-1}) \Longrightarrow 
\ti G = G^{-1}.
\label{b45}
\eea
For a diagonal metric $G= a^2 I$, and the invariance under the
scale factor duality  transformation (\ref{25}) is recovered as a particular
case of the global $O(d,d)$ symmetry of the low energy effective action. 

\section{The string cosmology equations}
\label{C}

In order to derive the cosmological equations let us include in the action,
for completeness, the antisymmetric tensor $B_{\mu\nu}$, a dilaton
potential $V(\phi)$, and also the possible contribution of other matter
sources represented by a Lagrangian density $L_m$:
\bea
&&
S= -{1\over 2\la_s^{d-1}} \int d^{d+1}x \sqrt{|g|}~
e^{-\phi}\left[R+ \left(\nabla \phi\right)^2- {1\over 12}
H_{\mu\nu\a}^2 + V(\phi)\right] \nonumber\\
&&
+\int d^{d+1}x \sqrt{|g|}~L_m. 
\label{c1}
\eea

In a scalar-tensor model of gravity, expecially in the presence of higher
derivative interactions, it is often convenient to write the action in the
language of exterior differential forms, as this may simplify the
variational procedure (see for instance \cite{86}). Here we will follow
however the more traditional approach, by varying the action with
respect to $g_{\mu\nu}, \phi$ and $B_{\mu\nu}$. We shall take into
account the dynamical stress tensor $T_{\mu\nu}$ of the matter sources
(defined in the usual way), as well as the scalar source $\sg$ representing
a possible direct coupling of the dilaton to the matter fields:
\beq
\da_g (\sqrt{-g} L_m) = {1\over 2} \sqrt{-g} T_{\mu\nu} \da g^{\mu\nu}, 
~~~~~~~~~~\da_\phi (\sqrt{-g} L_m) = \sqrt{-g} \sg \da \phi.
\label{c2}
\eeq

We start performing the variation with respect to the metric, using the
standard, general relativistic results:
\bea
&&
\da\sqrt{-g}= -{1\over 2} \sqrt{-g} g_{\mu\nu} \da g^{\mu\nu}, 
\nonumber \\
&&
\da \left ( \sqrt{-g} R\right) = \sqrt{-g}\left(G_{\mu\nu} \da g^{\mu\nu}
+g_{\mu\nu} \nabla^2 \da g^{\mu\nu}- \nabla_\mu \nabla_\nu 
\da g^{\mu\nu}\right), 
\label{c3}
\eea
where $G_{\mu\nu}$ is the usual Einstein tensor. It must be noted,
however, that the second covariant derivatives of $\da g^{\mu\nu}$,
when integrated by parts, are no longer equivalent to a divergence (and
then to a surface integral),  because of the dilaton factor $\exp(-\phi)$ in
front of the Einstein action, which adds dilatonic gradients to the full
variation. By performing a first integration by part, and using the metricity
condition $\nabla _\a g_{\mu\nu}=0$, we get in fact:
\bea
\fl \qquad
\da_g S = {1\over 2} \int d^{d+1}x \sqrt{|g|}T_{\mu\nu} \da g^{\mu\nu}
-{1\over 2\la_s^{d-1}} \int d^{d+1}x \sqrt{|g|}~
e^{-\phi}   \nonumber\\
\fl\qquad
\Bigg[G_{\mu\nu} +\nabla_\a \phi g_{\mu\nu} \nabla^\a 
-\nabla_\mu \phi \nabla_\nu +\nabla_\mu \phi \nabla_\nu \phi-
{1\over 2} g_{\mu\nu} \left(\nabla \phi\right)^2 
-{1\over 2} g_{\mu\nu} V(\phi)\nonumber\\
\fl\qquad
+ {1\over 2} g_{\mu\nu}{1\over 12}
H_{\a\b\ga}^2
 -{3\over 12}H_{\mu\a\b}H_\nu\,^{\a\b}\Bigg] \da
g^{\mu\nu} \nonumber\\
\fl\qquad
-{1\over 2\la_s^{d-1}} \int d^{d+1}x \sqrt{|g|}~\nabla_\a \Bigg[
e^{-\phi} g_{\mu\nu} \nabla^\a \da g^{\mu\nu}-e^{-\phi} \nabla_\nu 
\da g^{\nu\a}\Bigg]=0.
\label{c4}
\eea
A second integration by part of $\nabla \da g^{\mu\nu}$ cancels the
bilinear term $\nabla_\mu \phi \nabla_\nu \phi$, and leads to the field
equations:
\bea
\fl
G_{\mu\nu} + \nabla_\mu\nabla_\nu \phi +{1\over 2}g_{\mu\nu}
\left[\left(\nabla \phi\right)^2 -2 \nabla^2 \phi - V(\phi) +{1\over 12}
H_{\a\b\ga}^2\right] -{1\over 4} H_{\mu\a\b}H_\nu\,^{\a\b}
\nonumber\\
={1\over 2}e^\phi T_{\mu\nu}.
\label{c5}
\eea

We have chosen units such that $2 \la_s^{d-1}=1$, so that $e^\phi$
represents the ($d+1$)-dimensional gravitational constant (see Appendix
A). Also, we have implicitly added to the action the boundary term
\beq
{1\over 2\la_s^{d-1}} \int _{\pa \Om}\sqrt{|g|}~
e^{-\phi}K^\a d\Sigma_\a, 
\label{c5a}
\eeq
whose variation with respect to $g$ exactly cancels the contribution of the
total divergence appearing in the last integral of eq. (\ref{c4}):
\beq
\da_g \int \sqrt{|g|}~
e^{-\phi}K^\a d\Sigma_\a=  \int \sqrt{|g|}~
e^{-\phi}\left ( g_{\mu\nu} \nabla^\a \da g^{\mu\nu}- \nabla_\nu 
\da g^{\nu\a}\right)d\Sigma_\a.
\label{c5b}
\eeq
Here $K^\a$ is a geometric term representing the so-called extrinsic
curvature on the $d$-dimensional closed hypersurface, of infinitesimal
area $d\Sigma_\a$, bounding the total spacetime volume over which we
are varying the action. Note that the integral (\ref{c5a}) differs from the
usual boundary term, used in general relativity \cite{87} to derive the
Einstein equations, only by the presence of the tree-level dilaton coupling 
$e^{-\phi}$ to the extrinsic curvature. 

Let us now perform the variation with respect to the dilaton, again in
units  $2 \la_s^{d-1}=1$. We get the Eulero-Lagrange equations:
\bea
&&
\pa_\mu \left[ -2 \sqrt{-g} e^{-\phi}\pa^\mu\phi\right] =
e^{-\phi} \sqrt{-g}\left[R+ \left(\nabla \phi\right)^2- {1\over 12}
H^2 + V\right]
\nonumber\\
&&
- e^{-\phi} \sqrt{-g}V' + \sqrt{-g}\sg
\label{c6}
\eea
(where $V' =\pa V/\pa \phi$), from which
\beq
R+ 2 \nabla^2 \phi -\left(\nabla \phi\right)^2 +V-V'- {1\over 12} H^2+ 
e^\phi \sg =0.
\label{c7}
\eeq
The variation with respect to $B_{\mu\nu}$, 
\beq
\da_B  \int d^{d+1}x \sqrt{|g|}~
e^{-\phi}\left(\pa_\mu B_{\nu\a}\right)H^{\mu\nu\a} =0,
\label{c8}
\eeq
 gives finally 
\beq
\pa_\mu \left(\sqrt{|g|}~
e^{-\phi}H^{\mu\nu\a}\right) =0= \nabla_\mu \left
(e^{-\phi}H^{\mu\nu\a}\right).
\label{c9}
\eeq

Eqs. (\ref{c5}, \ref{c7}, \ref{c9}) are the equations governing the evolution
of the string cosmology background, at low energy. 
Note that eq. (\ref{c5}) can also be given in a simplified form: if we
eliminate the scalar curvature present inside the Einstein tensor, by using
the dilaton equation (\ref{c7}), we obtain:
\beq
R_\mu\,^\nu + \nabla_\mu\nabla^\nu \phi 
-{1\over 2} \da_\mu^\nu V' 
-{1\over 4}
H_{\mu\a\b}H^{\nu\a\b}={1\over 2} e^\phi \left(T_\mu\,^\nu-
\da_\mu^\nu \sg\right).
\label{c10}
\eeq

For the purpose of these lectures, it will be enough to derive some simple
solution of the string cosmology equations in the absence of the potential
($V=0$), of the antisymmetric tensor ($B=0$), and with a perfect fluid,
minimally coupled to the dilaton ($\sg=0$), as the matter sources.
Assuming for the background a Bianchi I type metric, we can work in the
syncronous gauge, by setting 
\bea
&&
g_{\mu\nu}= {\rm diag} (1, -a_i^2 \da_{ij}), ~~~~~~~
a_i=a_i(t),  ~~~~~~~\phi= \phi(t),
\nonumber\\
&&
T_\mu\,^\nu=  {\rm diag} (\rho, -p_i^2 \da_i^j), ~~~~~~~
p_i/\rho=\ga_i= {\rm const},  ~~~~~~~ \r=\r(t). 
\label{c11}
\eea
For this background:
\bea
&&
\Ga_{0i}\,^j=H_i\da_i^j, ~~~~ \Ga_{ij}\,^0= a_i \dot a_i 
\da_{ij},  ~~~~ R_0\,^0= - \sum_i \left(\dot H_i +H_i^2\right),
\nonumber\\
 &&
R_i\,^j= - \dot H_i\da_i^j -H_i\da_i^j \sum_kH_k,~~~~
R= -\sum_i\left( 2 \dot H_i + H_i^2\right)- \left(\sum_i H_i\right)^2, 
\nonumber\\
&&
\left(\nabla \phi\right)^2=\dot \phi^2, ~~~~\nabla^2\phi =\ddot \phi +
\sum_i H_i \dot \phi, ~~~~\nabla_0\nabla^0\phi= \ddot \phi,
\nonumber\\
&&
\nabla_i\nabla^j\phi= H_i \dot \phi \da_i^j.
 \label{c12}
\eea
The dilaton eq. (\ref{c7}) gives then
\beq
2 \ddot \phi +2 \dot\phi \sum_i H_i - \dot \phi ^2 
-\sum_i\left( 2 \dot H_i + H_i^2\right)- \left(\sum_i H_i\right)^2=0.
\label{c13}
\eeq
The $(00)$ component of the  eq. (\ref{c5}) gives 
\beq
\dot \phi^2 -2 \dot \phi\sum_iH_i - \sum_iH_i^2 +  
\left(\sum_i H_i\right)^2= e^\phi \r.
\label{c14}
\eeq
The diagonal, spatial components $(i,i)$ of eq. (\ref{c5}) (the off-diagonal
components are trivially satisfied) give
\bea
&&
\dot H_i + H_i \sum_k H_k -H_i\dot \phi -{1\over 2} 
\sum_i\left( 2 \dot H_i + H_i^2\right) -{1\over 2}\left(\sum_i H_i\right)^2
\nonumber \\
&&
-{1\over 2} \dot \phi^2 + \ddot \phi + \dot \phi\sum_i H_i
= {1\over 2} e^\phi p_i.
\label{c15}
\eea
The last five terms on the left-hand side add to zero because of the
dilaton equation (\ref{c13}), and the spatial equations reduce to
\beq
\dot H_i -H_i \left(\dot \phi - \sum_k H_k\right) =  {1\over 2} 
e^\phi p_i.
\label{c16}
\eeq

The above equations are clearly invariant under time-reversal, $t \ra -t$.
In order to make explicit also their duality invariance, let us introduce
again the shifted dilaton (see eq. (\ref{b7})), such that
\beq
e^{\fb} = e^\phi/ \sqrt{-g}, ~~~~~~~~~~~~
\fbp = \dot \phi - \sum_iH_i,
\label{c17}
\eeq
and define
\beq
\rb=  \r \sqrt{-g}= \r \prod_i a_i, ~~~~~~~~~~~~~
\pb= p \sqrt{-g}= p\prod_i a_i.
\label{c18}
\eeq
In terms of these variables, the time and space equations
(\ref{c14},\ref{c16}), and the dilaton equation (\ref{c13}), become,
respectively: 
\bea
&&
\fbp^2 -\sum_iH_i^2= e^{\fb} \rb, \label{c19}\\
&&
\dot H_i -H_i \fbp={1\over 2} e^{\fb} \pb_i, \label{c20}\\
&&
2 \ddot{\fb} -\fbp^2 -\sum_iH_i^2=0.
\label{c21}
\eea
They are explicitly invariant under the scale-factor duality transformation:
\beq
a_i \ra a_i^{-1}, ~~~~~~~ \fb \ra \fb, ~~~~~~~
\rb \ra \rb, ~~~~~~~ \pb \ra -\pb.
\label{c21a}
\eeq
which implies, for a perfect fluid source, a ``reflection" of the equation of
state, $\ga =p/\r =\pb/\rb \ra -\pb/\rb = -\ga$ (see the first paper in Ref.
\cite{8}). A general $O(d,d)$ transformation changes however the equation
of state in a more drastic way (see \cite{12}), introducing also shear and
bulk viscosity. 

The above $(d+2)$ equations are a system of independent equations for the 
$(d+2)$ variables $\{a_i, \phi, \r\}$. Their combination implies the usual
covariant conservation  of the energy density. By
differentiating eq. (\ref{c19}), and using (\ref{c20},\ref{c21}) to eliminate
$\dot H_i$, $\fbp$, respectively, we get in fact
\beq
\dot {\rb} +\sum_i H_i \pb_i=0,
\label{c22}
\eeq
which, using the definitions (\ref{c18}), is equivalent to
\beq
\dot \r + \sum_iH_i \left(\r +p_i\right)=0.
\label{c23}
\eeq
In order to obtain exact solutions, it is convenient to include  this
energy conservation equation in the full system of independent equations. 

In these lectures I will present only a particular example of
matter-dominated solution by considering a $d$-dimensional, isotropic
background characterized by a power-law evolution,
\beq
a \sim t^\a, ~~~~~~~~~ \fb \sim -\b \ln t, ~~~~~~~~~p=\ga \r. 
\label{c24}
\eeq
We use (\ref{c19},\ref{c21}, \ref{c22}) as independent equations. The
integration of eq. (\ref{c22}) gives immediately
\beq
\rb =\r_0a^{-d\ga};
\label{c25}
\eeq
eq. (\ref{c19}) is then satisfied provided
\beq
d\ga \a +\b=2.
\label{c26}
\eeq
Finally, eq. (\ref{c21}) provides the constraint
\beq
2\b -\b^2 -d \a^2 =0.
\label{c27}
\eeq
We have then a system of two equations for the two parameters 
$\a,\b$ (note that, if $\a$ is a solution for a given $\ga$, then also $-\a$ is
a solution, associated to $-\ga$). We have in general two solutions. The
trivial flat space solution  $\b=2, \a=0$, corresponds to dust matter
($\ga=0$) according to eq. (\ref{c20}). For $\ga \not= 0$ we obtain instead
\beq 
\a = {2 \ga\over 1+ d\ga^2}, ~~~~~~~~~~~
\b = {2 \over 1+ d\ga^2},
\label{c28}
\eeq
which fixes the time evolution of $a$ and $\fb$:
\beq
a \sim t^{2\ga\over 1+ d\ga^2}, ~~~~~~~~~~
\fb = - {2 \over 1+ d\ga^2}\ln t,
\label{c29}
\eeq
and also of the more conventional variables $\r, \phi$:
\beq
\r = \rb a^{-d} = \r_0 a^{-d(1+\ga)}, ~~~~~~~~~~
\phi = \fb +d \ln a=  {2(d\ga -1) \over 1+ d\ga^2}\ln t.
\label{c30}
\eeq

This particular solution reproduces the small curvature limit of  the 
general  solution with perfect fluid sources (see the last two papers of
Ref. \cite{8}), sufficiently far from the  singularity.  Like in the
vacuum solution (\ref{b20}) there are four branches, related by
time-reversal and by the duality transformation (\ref{c21a}), and
characterized by the scale factors
\beq
a_{\pm} (\pm t) \sim (\pm t) ^{\pm 2 \ga/(1+d\ga^2)}.
\label{c31}
\eeq
The duality transformation that preserves $\fb$ and $\rb$, and inverts the
scale factor, in this case is simply represented by the
transformation $\ga \ra -\ga$. Consider for instance the standard
radiation-dominated solution, corresponding to $d=3$, $\ga=1/3$, and
$t>0$, and associated to a constant dilaton, according to eq. (\ref{c30}). A
duality transformation gives a new solution with $\ga=-1/3$, namely
(from (\ref{c29},\ref{c30}):
\beq
a \sim t^{-1/2}, ~~~~~~~~\r \sim a^{-2}, ~~~~~~~~
\phi \sim -3 \ln t.
\label{c32}
\eeq
By performing an additional time reflection we then obtain the pre-big
bang solution ``dual to radiation",  already reported in eq. (\ref{216}).

\end{document}